\documentclass{LMCS}

\def\dOi{10(3:11)2014}
\lmcsheading%
{\dOi}
{1--32}
{}
{}
{Nov.~12, 2013}
{Aug.~26, 2014}
{}

\ACMCCS{[{\bf Mathematics of computing}]: Markov processes; [{\bf
      Theory of computation}]: Probabilistic computation}

\usepackage[utf8]{inputenc}
\usepackage{cite}
\usepackage[T1]{fontenc}
\usepackage{hyperref}
\usepackage{amsfonts}
\usepackage{multirow}
\usepackage{amsmath,amssymb,ifthen,tabularx,dsfont,balance,subfig,wrapfig,url,graphicx,color}

\makeatletter
\def\Hv@scale{.85}
\renewcommand*\subjclass[2][1991]{%
  \def\@subjclass{#2}%
  \@ifundefined{subjclassname@#1}{%
    \ClassWarning{\@classname}{Unknown edition (#1) of ACM
      Subject Classification; using '1991'.}%
  }{%
    \@xp\let\@xp\subjclassname\csname subjclassname@#1\endcsname
  }%
}
\@namedef{subjclassname@2012}{2012 ACM Subject Classification}
\makeatother

\usepackage{mysty}

\usepackage{microtype}

\begin{document}

\title{Refinement and Difference for Probabilistic Automata}

\author[B.~Delahaye]{Beno\^{\i}t Delahaye\rsuper a}
\address{{\lsuper a}Universit{\'e} de Nantes, France}
\email{benoit.delahaye@univ-nantes.fr}

\author[U.~Fahrenberg]{Uli Fahrenberg\rsuper b}
\address{{\lsuper{b,d}}Inria / IRISA Rennes, France}
\email{\{ulrich.fahrenberg,axel.legay\}@irisa.fr}

\author[K.~G.~Larsen]{Kim G.\ Larsen\rsuper c}
\address{{\lsuper c}Aalborg University, Denmark}
\email{kgl@cs.aau.dk}

\author[A.~Legay]{Axel Legay\rsuper d}
\address{\vspace{-18 pt}}


\titlecomment{This is an extended version of the
  paper~\cite{DBLP:conf/qest/DelahayeFLL13} which has been presented at
  the 10th International Conference on Quantitative Evaluation of
  SysTems (QEST 2013) in Buenos Aires, Argentina.  Compared
  to~\cite{DBLP:conf/qest/DelahayeFLL13}, and in addition to a number of
  small changes and improvements, proofs of the main statements and a
  new section on counter-example generation have been added to the
  paper.}

\begin{abstract}
  This paper studies a difference operator for stochastic systems whose
  specifications are represented by Abstract Probabilistic Automata
  (APAs). In the case refinement fails between two specifications, the
  target of this operator is to produce a specification APA that
  represents all witness PAs of this failure.  Our contribution is an
  algorithm that permits to approximate the difference of two
  deterministic APAs with arbitrary precision. Our technique relies on
  new quantitative notions of distances between APAs used to assess
  convergence of the approximations, as well as on an in-depth
  inspection of the refinement relation for APAs. The procedure is
  effective and not more complex than refinement checking.
\end{abstract}

\maketitle

\section{Introduction}\label{sec:intro}

Probabilistic automata as promoted by Segala and
Lynch~\cite{DBLP:conf/concur/SegalaL94} are a widely-used formalism for
modeling systems with probabilistic behavior.  These include randomized
security and communication protocols, distributed systems, biological
processes and many other applications.  Probabilistic model
checking~\cite{DBLP:conf/tacas/HintonKNP06, books/BaierK08,
  DBLP:conf/focs/Vardi85} is then used to analyze and verify the
behavior of such systems.  Given the prevalence of applications of such
systems, probabilistic model checking is a field of great interest.
However, and similarly to the situation for non-probabilistic model
checking,
probabilistic model checking suffers from \emph{state space explosion},
which hinders its applicability considerably.

One generally successful technique for combating state space explosion
is the use of \emph{compositional} techniques, where a (probabilistic)
system is model checked by verifying its components one by one.  This
compositionality can be obtained by \emph{decomposition}, that is, to
check whether a given system satisfies a property, the system is
automatically decomposed into components which are then verified.
Several attempts at such automatic decomposition techniques have been
made~\cite{DBLP:conf/tacas/CobleighGP03,DBLP:conf/tacas/KwiatkowskaNPQ10},
but in general, this approach has not been very
successful~\cite{DBLP:journals/tosem/CobleighAC08}.

As an alternative to the standard model checking approaches using
logical specifications, e.g.~LTL, MITL or PCTL~\cite{books/MannaP92,
  DBLP:journals/jacm/AlurFH96, DBLP:journals/fac/HanssonJ94},
automata-based specification theories have been proposed, such as
Input/Output Automata~\cite{Lynch-tuttle88}, Interface
Automata~\cite{DBLP:conf/sigsoft/AlfaroH01}, and Modal
Specifications~\cite{DBLP:conf/avmfss/Larsen89, TheseJBR07,
  DBLP:conf/csr/BauerFLT12}.  These support composition \emph{at
  specification level}; hence a model which naturally consists of a
composition of several components can be verified by model checking each
component on its own, against its own specification.  The overall model
will then automatically satisfy the composition of the component
specifications.  Observe that this solves the decomposition problem
mentioned above: instead of trying to automatically decompose a system
for verification, specification theories make it possible to verify the
system without constructing it in the first place.

Moreover, specification theories naturally support \emph{stepwise
  refinement} of specifications, i.e.~iterative implementation of
specifications, and \emph{quotient}, i.e.~the synthesis of missing
component specifications given an overall specification and a partial
implementation.  Hence they allow both logical and compositional
reasoning at the same time, which makes them well-suited for
compositional verification.

For probabilistic systems, such automata-based specification theories
have been first introduced in~\cite{DBLP:conf/lics/JonssonL91}, in the
form of Interval Markov Chains.  The focus there is only on refinement
however; to be able to consider also composition and conjunction, we
have in~\cite{TCS11} proposed Constraint Markov Chains (CMCs) as a
natural generalization which uses general constraints instead of
intervals for next-state probabilities.

In~\cite{VMCAI11}, we have extended this specification theory to
probabilistic automata, which combine stochastic and non-deterministic
behaviors.  These \emph{Abstract Probabilistic Automata} (APA) combine
modal specifications and CMCs.  Our specification theory using APA
should be viewed as an alternative to classical
PCTL~\cite{DBLP:journals/fac/HanssonJ94}, probabilistic I/O
automata~\cite{DBLP:books/mk/Lynch96} and stochastic extensions of
CSP~\cite{DBLP:journals/tcs/HermannsHK02}.  Like these, its purpose is
model checking of probabilistic properties, but unlike the alternatives,
APA support compositionality at specification level.

In the context of refinement of specifications, it is important that
informative debugging information is given in case refinement fails.
More concretely, given APAs $N_1$, $N_2$ for which $N_1$ does not refine
$N_2$, we would like to know \emph{why} refinement fails, and if
possible, \emph{where} in the state spaces of $N_1$ and $N_2$ there is a
problem.  We hence need to be able to compare APAs at the semantic
level, i.e.~to capture the \emph{difference between their sets of
  implementations} and to relate it to structural differences of the
APAs.  This is what we attempt in this paper: given two APAs $N_1$ and
$N_2$, to generate another APA $N$ such that the set of implementations
of $N$ is the differences between the sets of implementations of $N_1$
and of $N_2$.

As a second contribution, we introduce a notion of \emph{distance}
between APAs which measures how far away one APA is from refining a
second one.  This distance, adapted from our work
in~\cite{DBLP:conf/fsttcs/FahrenbergLT11, DBLP:conf/csr/BauerFLT12}, is
\emph{accumulating} and \emph{discounted}, so that differences between
APAs accumulate along executions, but in a way so that differences
further in the future are discounted, i.e.~have less influence on the
result than had they occurred earlier.

Both difference and distances are important tools to compare APAs which
are not in refinement.  During an iterative development process, one
usually wishes to successively replace specifications by more refined
ones, but due to external circumstances such as, for example, cost of
implementation, it may happen that a specification needs to be replaced
by one which is not a refinement of the old one.  This is especially
important when models incorporate quantitative information, such as for
APAs; the reason for the failed refinement might simply be some changes
in probability constraints, for example due to measurement updates.  In
this case, it is important to assess precisely \emph{how much} the new
specification differs from the old one.  Both the distance between the
new and old specifications, as well as their precise difference, can aid
in this assessment.

Unfortunately, because APAs are finite-state structures, the difference
between two APAs cannot always itself be represented by an APA.  Instead
of extending the formalism, we propose to \emph{approximate} the
difference for a subclass of APAs.  We introduce both over- and
under-approximations of the difference of two \emph{deterministic} APAs.
We construct a sequence of under-approximations which converges to the
exact difference, hence eventually capturing all PAs in
$\impl{N_1}\setminus \impl{ N_2}$, and a fixed over-approximation which
may capture also PAs which are not in the exact difference, but whose
distance to the exact difference is zero: hence any superfluous PAs
which are captured by the over-approximation are infinitesimally close
to the real difference.  Taken together, these approximations hence
solve the problem of assessing the precise difference between
deterministic APAs in case of failing refinement.

For completeness, we show as a last contribution how our algorithms can
be refined into a procedure that computes a single counter-example to a
failed refinement.

We restrict ourselves to the subclass of deterministic APAs, as it
permits syntactic reasoning to decide and compute refinement. Indeed,
for deterministic APAs, syntactic refinement coincides with semantic
refinement~\cite{VMCAI11}, hence allowing for efficient procedures. Note
that although the class of APAs we consider is called ``deterministic'',
it still offers non-determinism in the sense that one can choose between
different actions in a given state.

\subsection*{Related work}

This paper embeds into a series of articles on APA as a specification
theory~\cite{VMCAI11, DBLP:conf/acsd/DelahayeKLLPSW11, QEST11,
  DBLP:conf/qest/DelahayeFLL13, journals/iandc/DelahayeKLLPSW13}.
In~\cite{VMCAI11} we introduce deterministic APA, generalizing earlier
work on interval-based abstractions of probabilistic
systems~\cite{FLW06,DBLP:conf/lics/JonssonL91, KKLW07}, and define
notions of refinement, logical composition, and structural composition.
We also introduce a notion of \emph{compositional abstraction} for APA.
In~\cite{DBLP:conf/acsd/DelahayeKLLPSW11} we extend this setting to
non-deterministic APA and give a notion of (over-approximating)
determinization.  In~\cite{QEST11} we introduce the tool APAC which
implements most of these operations and hence can be used for
compositional design and verification of probabilistic systems.

The journal paper~\cite{journals/iandc/DelahayeKLLPSW13} sums up and
streamlines the contributions of~\cite{VMCAI11,
  DBLP:conf/acsd/DelahayeKLLPSW11, QEST11}.  One interesting detail in
the theory of APA is that there are several types of \emph{syntactic
  refinement} of APA.  In~\cite{journals/iandc/DelahayeKLLPSW13}, these
are called \emph{strong} refinement, \emph{weak} refinement, and
\emph{weak weak} refinement, respectively; all are motivated by similar
notions for CMCs~\cite{TCS11}.  For \emph{deterministic} APAs, these
refinements agree, and they also coincide with thorough refinement
(i.e.~inclusion of implementation sets).  The distance and difference we
introduce in the present paper complement the refinement and abstraction
from~\cite{journals/iandc/DelahayeKLLPSW13}, in the sense that our
distance between APAs is a quantitative generalization of APA
refinement, and our difference structurally characterizes refinement
failure.

Compositional abstraction of APA is also considered
in~\cite{DBLP:conf/ifipTCS/SherK12}, but with the additional feature
that transitions with the same action (i.e.~non-deterministic choices)
can be combined into so-called \emph{multi-transitions}.  The refinement
in~\cite{DBLP:conf/ifipTCS/SherK12} is thus even weaker than the weak
weak refinement of~\cite{journals/iandc/DelahayeKLLPSW13}; for
deterministic APA however, they agree.

Differences between automata-based specifications have not been
considered much in the literature.
\cite{DBLP:journals/sosym/SassolasCU11}~develops a notion of
\emph{pseudo-merge} between modal specifications which keeps track of
inconsistencies between specifications; here, the inconsistent states
can be seen as a form of difference.  Distances between probabilistic
systems have been introduced in~\cite{AMRS07,
  DBLP:journals/tcs/DesharnaisGJP04, DBLP:conf/fossacs/BreugelMOW03} and
other works, and distances between modal specifications
in~\cite{DBLP:conf/csr/BauerFLT12, DBLP:journals/fmsd/BauerFJLLT13,
  DBLP:conf/mfcs/BauerFJLLT11}; here, we combine these notions to
introduce distances between APAs.

The originality of our present work is the ability to measure how far
away one probabilistic specification is from being a refinement of
another, using distances and our new difference operator.  Both are
important in assessing precisely how much one APA differs from another.

\subsection*{Acknowledgment}

The authors wish to thank Joost-Pieter Katoen for interesting discussions
and insightful comments on the subject of this work, and a number of
anonymous referees for useful comments and improvements.

\section{Background}
\label{sec:preliminaries}
\label{sec:back}

Let $\Dist(S)$ denote the set of all discrete probability distributions
over a finite set $S$ and $\mathbb{B}_{2}=\{\top,\bot\}$.

\begin{defi}\label{def:probabilisticautomata}
  A probabilistic automaton (PA)~\cite{DBLP:conf/concur/SegalaL94} is a
  tuple $(S,A,L,AP,V,s_{0})$, where $S$ is a finite set of states with
  the initial state $s_{0} \in S$, $A$ is a finite set of actions, $L$:
  $S\times A\times \Dist(S)\rightarrow \mathbb{B}_{2}$ is a (two-valued)
  transition function, $AP$ is a finite set of atomic propositions and
  $V$: $S \rightarrow 2^{AP}$ is a state-labeling function.
\end{defi}

Consider a state $s$, an action $a$, and a probability distribution
$\mu$. The value of $L(s,a,\mu)$ is set to $\top$ in case there exists a
transition from $s$ under action $a$ to a distribution $\mu$ on
successor states. In other cases, we have $L(s,a,\mu)=\bot$.  We now
introduce Abstract Probabilistic Automata (APA)~\cite{VMCAI11}, that is
a specification theory for PAs. For a finite set $S$, we let $C(S)$
denote the set of constraints over discrete probability distributions on
$S$. Each element $\phi \in C(S)$ describes a set of distributions:
$Sat(\phi)\subseteq \Dist(S)$. Let
$\mathbb{B}_{3}=\{\top,\may,\bot\}$. APAs are formally defined as follows.

\begin{defi}\label{def:abstractprobabilisticautomata}
  An APA~\cite{VMCAI11} is a tuple $(S,A,L,AP,V,S_{0})$, where $S$ is a
  finite set of states, $S_0\subseteq S$ is a set of initial states, $A$
  is a finite set of actions, and $AP$ is a finite set of atomic
  propositions.  $L:S\times A\times C(S) \rightarrow \mathbb{B}_{3}$ is
  a \emph{three}-valued distribution-constraint function, and $V:S\!
  \rightarrow\!  2^{2^{AP}}$ maps each state in $S$ to a set of
  admissible labelings.
\end{defi}

APAs play the role of specifications in our framework. An APA transition
abstracts transitions of certain unknown PAs, called its
implementations.  Given a state $s$, an action $a$, and a constraint
$\phi$, the value of $L(s,a,\phi)$ gives the modality of the transition.
More precisely, the value $\top$ means that transitions under $a$ must
exist in the PA to some distribution in $Sat(\phi)$; $\may$ means that
these transitions are allowed to exist; $\bot$ means that such
transitions must not exist.  We will sometimes view $L$ as a
\emph{partial} function, with the convention that a lack of value for a
given argument is equivalent to the $\bot$ value.  The function $V$
labels each state with a subset of the power set of $AP$, which models a
disjunctive choice of possible combinations of atomic propositions. 

We say that an APA $N = (S,A,L,AP,V,S_{0})$ is in \emph{Single Valuation
  Normal Form} (SVNF) if the valuation function $V$ assigns at most one
valuation to all states, i.e.~$\forall s \in S, |V(s)| \le
1$. From~\cite{VMCAI11}, we know that every APA can be turned into an
APA in SVNF with the same set of implementations. An APA is
\emph{deterministic}~\cite{VMCAI11} if (1) there is at most one outgoing
transition for each action in all states, (2) two states with
overlapping atomic propositions can never be reached with the same
transition, and (3) there is only one initial state.

Note that every PA is an APA in SVNF where all constraints represent a
single distribution. As a consequence, all the definitions we present
for APAs in the following can be directly extended to PAs.

Let $N=(S,A,L,AP,V,\{s_0\})$ be an APA in SVNF and let $v \subseteq AP$.
Given a state $s \in S$ and an action $a \in A$, we will use the
notation $\textsf{succ}_{s,a}(v)$ to represent the set of potential
$a$-successors of $s$ that have $v$ as their valuation. Formally,
$\textsf{succ}_{s,a}(v) = \{s' \in S \st V(s') = \{v\}, \exists \phi \in
C(S), \mu \in Sat(\phi): L(s,a,\phi) \ne \bot, \mu(s')>0\}$.
When clear from the context, we may 
use $\textsf{succ}_{s,a}(s')$ instead of
$\textsf{succ}_{s,a}(V(s'))$. Observe that when $N$ is deterministic, we
have $|\textsf{succ}_{s,a}(v)| \le 1$ for all $s,a,v$.

\section{Refinement and Distances between APAs}
\label{sec:ref+dist}

We recall the notion of refinement between APAs. Roughly speaking,
refinement guarantees that if $A_1$ refines $A_2$, then the set of
implementations of $A_1$ is included in the one of $A_2$. 

\begin{defi}
  \label{def:leqbox}
  Let $S$ and $S'$ be non-empty sets and $\mu \in \Dist(S)$, $\mu' \in
  \Dist(S')$.  We say that $\mu$ is \emph{simulated} by $\mu'$ with
  respect to a relation $\rel \subseteq S\times S'$ and a
  \emph{correspondence function} $\delta: S \to (S'\to[
  0,1])$~\cite{VMCAI11} if
  \begin{enumerate}
  \item for all $s\in S$ with $\mu(s)>0$, $\delta(s)$ is a distribution
    on $S'$,
  \item for all $s'\in S'$, $\sum_{s\in S} \mu(s)\cdot
    \delta(s)(s')=\mu'(s')$, and
  \item \label{it:distsim-3} whenever $\delta(s)(s')>0$, then $(s, s')
    \in \rel$.
  \end{enumerate}
\end{defi}

We write $\mu \leqbox_{\rel}^{\delta} \mu'$ if $\mu$ is simulated by
$\mu'$ with respect to $\rel$ and $\delta$, $\mu \leqbox_{\rel} \mu'$ if
there exists $\delta$ with $\mu \leqbox_{\rel}^{\delta} \mu'$, and
$\mu\leqbox^\delta \mu'$ for $\mu\leqbox_{S \times S'}^\delta \mu'$.


\begin{defi}
  \label{apaweakweakrefdef}\label{def:refinement}
  Let $N_1=(S_1,A,L_1,AP,V_1,S_0^1)$ and $N_2=(S_2,A, L_2,AP,V_2,S_0^2)$
  be APAs.  A relation $\rel\subseteq S_1 \times S_2$ is a
  \emph{refinement} relation~\cite{VMCAI11} if, for all $(s_1,s_2)\in
  \rel$, we have $V_1(s_1) \subseteq V_2(s_2)$ and
  \begin{enumerate}
  \item $\forall a \in A,\, \forall \phi_2 \in C(S_2)$, if
    $L_2(s_2,a,\phi_2) = \top$, then $\exists \phi_1\in C(S_1):
    L_1(s_1,a,\phi_1) = \top$ and $\forall \mu_1 \in Sat(\phi_1),\,
    \exists\mu_2 \in Sat(\phi_2)$ such that $\mu_1 \leqbox_{\rel}
    \mu_2$,
  \item $\forall a\in A,\, \forall \phi_1 \in C(S_1)$, if
    $L_1(s_1,a,\phi_1) \neq \bot$, then $\exists \phi_2 \in C(S_2)$ such
    that $L_2(s_2,a,\phi_2) \ne \bot$ and $\forall \mu_1 \in
    Sat(\phi_1)$, $\exists\mu_2 \in Sat(\phi_2)$ such that $\mu_1
    \leqbox_{\rel} \mu_2$.
  \end{enumerate}
\end{defi}

\noindent We say that $N_1$ refines $N_2$, denoted $N_1 \preceq N_2$, if there
exists a refinement relation such that $\forall s_0^1 \in S_0^1: \exists
s_0^2 \in S_0^2: (s_0^1, s_0^2) \in \rel$. Since any PA $P$ is also an
APA, we say that $P$ \emph{satisfies} $N$ (or equivalently $P$
\emph{implements} $N$), denoted $P\models N$, if $P \preceq N$.  In the
following, a refinement relation between a PA and an APA is called a
{\em satisfaction} relation. In~\cite{VMCAI11}, it is shown that for
deterministic APAs $N_1$, $N_2$, we have $N_1 \preceq N_2 \iff
\impl{N_1} \subseteq \impl{N_2}$, where $\impl{N_i}$ denotes the set of
implementations of APA $N_i$.  Hence for deterministic APAs, the
difference $\impl{ N_1}\setminus \impl{ N_2}$ is non-empty iff
$N_1\not\preceq N_2$.  This equivalence breaks for non-deterministic
APAs~\cite{VMCAI11}, whence we develop our theory only for deterministic
APAs.

To show a convergence theorem about our difference construction in
Sect.~\ref{sec:under-approx} below, we need a relaxed notion of
refinement which takes into account that APAs are a \emph{quantitative}
formalism.  Indeed, refinement as of Def.~\ref{def:refinement} is a
purely qualitative relation; if both $N_2\not\preceq N_1$ and
$N_3\not\preceq N_1$, then there are no criteria to compare $N_2$ and
$N_3$ with respect to $N_1$, saying which one is the closest to $N_1$.
We provide such a relaxed notion by generalizing refinement to a
\emph{discounted distance} which provides precisely such criteria. In
Sect.~\ref{sec:under-approx}, we will show how those distances can be
used to show that increasingly precise difference approximations between
APAs converge to the real difference.

In order to simplify notation, the definitions presented below are
dedicated to APAs in SVNF.  They can however be easily extended to
account for general APAs.  The next definition shows how a distance
between states is lifted to a distance between constraints.

\begin{defi}
  Let $d : S_1 \times S_2 \rightarrow \mathbb{R}^{+}$ and $\phi_1 \in
  C(S_1)$, $\phi_2 \in C(S_2)$ be constraints in $N_1$ and $N_2$. Define
  the distance $D_{N_1,N_2}$ between $\phi_1$ and $\phi_2$ as follows:
  \begin{equation*}
    D_{N_1,N_2} (\phi_1,\phi_2,d) = \sup_{\mu_1 \in Sat(\phi_1)\,}
    \inf_{ \mu_2\in Sat( \phi_2)\,} \inf_{ \delta:
      \mu_1\leqbox^\delta \mu_2} \sum_{( s_1,s_2) \in S_1\times S_2}
    \mu_1(s_1)\delta(s_1)( s_2) d(s_1,s_2) 
  \end{equation*}
\end{defi}\medskip

\noindent Note the analogy of this definition to the one of the \emph{Hausdorff}
distance between (closed) subsets of a metric space: Any distribution
$\mu_1$ in $Sat( \phi_1)$ is sought matched with a distribution $\mu_2$
in $Sat( \phi_2)$ which mimics it as closely as possible, where the
quality of a match is measured by existence of a correspondence function
$\delta$ which minimizes the distance between points reached from $s_1$
and $s_2$ weighted by their probability.

For the definition of $d$ below, we say that states $s_1 \in S_1$, $s_2
\in S_2$ are \emph{not compatible} if
\begin{enumerate}
\item $V_1(s_1) \ne V_2(s_2)$,
\item there exists $a \in A$ and $\phi_1 \in C(S_1)$ such that
  $L_1(s_1,a,\phi_1) \ne \bot$ and for all $\phi_2 \in C(S_2),
  L_2(s_2,a,\phi_2) = \bot$, or
\item there exists $a \in A$ and $\phi_2 \in C(S_2)$ such that
  $L_2(s_2,a,\phi_2) = \top$ and for all $\phi_1 \in C(S_1),
  L_1(s_1,a,\phi_1) \ne \top$.
\end{enumerate}
For compatible states, their distance is similar to the accumulating
branching distance on modal transition systems as introduced
in~\cite{DBLP:conf/csr/BauerFLT12, DBLP:conf/fsttcs/FahrenbergLT11},
adapted to our formalism. In the rest of the paper, the real constant $0
< \lambda < 1$ represents a discount factor.  Formally, $d : S_1 \times
S_2 \rightarrow [0,1]$ is the least fixed point to the following system
of equations:
\begin{equation}
  \label{eq:lfpdist}
  d(s_1,s_2) =
  \begin{cases}
    1 \text{ if } s_1 \text{ is not compatible with } s_2 \\
    \max
    \begin{cases}
      \displaystyle %
      \max_{ a,\phi_1: L_1(s_1,a,\phi_1) \ne \bot\,} \min_{ \phi_2:
        L_2(s_2,a,\phi_2) \ne \bot} \lambda
      D_{N_1,N_2}(\phi_1,\phi_2,d) \\
      \displaystyle %
      \max_{ a,\phi_2: L_2(s_2,a,\phi_2) = \top\,} \min_{ \phi_1:
        L_1(s_1,a,\phi_1) = \top} \lambda D_{N_1,N_2}(\phi_1,\phi_2,d)
    \end{cases}
    \text{\!\!\!\!otherwise}
  \end{cases}
\end{equation}
Since the above system of linear equations defines a \emph{contraction},
the existence and uniqueness of its least fixed point is ensured,
cf.~\cite{DBLP:journals/tcs/LarsenFT11}.
The intuition here is that $d( s_1, s_2)$ compares not only the
probability constraints at $s_1$ and $s_2$, but also (recursively) the
constraints at all states reachable from $s_1$ and $s_2$, weighted by
their probability.  Each step is discounted by $\lambda$, hence steps
further in the future contribute less to the distance.

The above definition intuitively extends to PAs, which allows us to
propose the two following notions of distance:

\begin{defi}
  \label{def:distances}
  Let $N_1 = (S_1,A,L_1,AP,V_1, S_0^1)$ and $N_2 = (S_2,A,L_2,AP,V_2,
  S_0^2)$ be APAs in SVNF. The \emph{syntactic} and \emph{thorough}
  distances between $N_1$ and $N_2$ are defined as follows:
  \begin{itemize}
  \item syntactic distance: $d(N_1,N_2) = \max_{s_0^1 \in S_0^1}
    \big(\min_{s_0^2 \in S_0^2} d(s_0^1, s_0^2) \big)$.
  \item thorough distance: $d_t(N_1,N_2) = \sup_{P_1 \in \sem{N_1}}
    \big( \inf_{P_2 \in \sem{N_2}} d(P_1,P_2) \big)$.
  \end{itemize}
\end{defi}\medskip

\noindent Note that the notion of thorough distance defined above intuitively
extends to sets of PAs: given two sets of PAs $\mathbb{S}_{1}$,
$\mathbb{S}_{2}$, we have $d_t(\mathbb{S}_1,\mathbb{S}_2) = \sup_{P_1
  \in \mathbb{S}_1} \big( \inf_{P_2 \in \mathbb{S}_2} d(P_1,P_2) \big)$.

We also remark that $N_1\preceq N_2$ implies $d( N_1, N_2)= 0$.  It can
be shown, cf.~\cite{DBLP:journals/jlp/ThraneFL10}, that both $d$ and
$d_t$ are \emph{asymmetric pseudometrics} (or \emph{hemimetrics}),
i.e.~satisfying $d( N_1, N_1)= 0$ and $d( N_1, N_2)+ d( N_2, N_3)\ge d(
N_1, N_3)$ for all APAs $N_1, N_2, N_3$ (and similarly for $d_t$).  The
fact that they are only pseudometrics, i.e.~that $d( N_1, N_2)= 0$ does
not imply $N_1= N_2$, will play a role in our convergence arguments
later.

The following proposition shows that the thorough distance is bounded
above by the syntactic distance.  Hence we can bound distances between
(sets of) implementations by the syntactic distance between their
specifications.

\begin{prop}
\label{th:distances}
For all APAs $N_1$ and $N_2$ in SVNF, it holds that $d_t(N_1,N_2) \le
d(N_1,N_2)$.
\end{prop}

\proof
  For a distribution $\mu_1$ and a constraint $\phi_2$, we denote by
  \begin{equation*}
    \rd( \mu_1, \phi_2):=\{ \delta: \mu_1\leqbox^\delta \mu_2\mid \mu_2\in
    Sat( \phi_2)\}
  \end{equation*}
  the set of all correspondence functions between $\mu_1$ and
  distributions satisfying $\phi_2$.

  If $d( N_1, N_2)= 1$, we have nothing to prove.  Otherwise, write
  $N_i=( S_i, A, L_i, AP, V_i, S_0^i)$ for $i= 1, 2$, and let $P_1=(
  S_1', A, L_1', AP, V_1', \bar S_0^1)\in \sem{ N_1}$ and $\eta> 0$; we
  need to expose $P_2\in \sem{ N_2}$ for which $d( P_1, P_2)\le d( N_1,
  N_2)+ \eta$.  Note that by the triangle inequality, $d( P_1, N_2)\le
  d( P_1, N_1)+ d( N_1, N_2)\le d( N_1, N_2)$.  Define $P_2=( S_2, A,
  L_2', AP, V_2, S_0^2)$, with $L_2'$ given as follows:

  For all $s_1'\in S_1'$, $a\in A$, $\mu_1\in Dist( S_1')$ for which
  $L_1'( s_1', a, \mu_1)= \top$ and for all $s_2\in S_2$, $\epsilon< 1$
  with $\epsilon:= d( s_1', s_2)< 1$: We must have $\phi_2\in Dist(
  S_2)$ such that $L_2( s_2, a, \phi_2)\ne \bot$ and
  \begin{equation*}
    \inf_{\delta \in
      \rd(\mu_1,\phi_2)} \sum_{( t_1', t_2)\in S_1'\times S_2}
    \mu_1(t_1')\delta(t_1',t_2) d(t_1',t_2)\le \lambda^{ -1} \epsilon\,,
  \end{equation*}
  so there must exist a correspondence function $\delta \in
  \rd(\mu_1,\phi_2)$ for which
  \begin{equation*}
    \sum_{( t_1', t_2)\in
      S_1'\times S_2} \mu_1(t_1')\delta(t_1',t_2) d(t_1',t_2)\le \lambda^{
      -1} \epsilon+ \lambda^{ -1} \eta.
  \end{equation*}
  We let $\mu_2( s)= \sum_{s_1' \in S_1} \mu_1(s_1') \delta(s_1',s)$ and
  set $L_2'( s_2, a, \mu_2)= \top$ in $P_2$.

  Similarly, for all $s_2\in S_2$, $a\in A$, $\phi_2\in C( S_2)$ for
  which $L_2( s_2, a, \phi_2)= \top$ and for all $s_1'\in S_1'$ with
  $\epsilon:= d( s_1', s_2)< 1$: We must have $\mu_1\in Dist( S_1')$ for
  which $L_1'( s_1', a, \mu_1)= \top$ and
  \begin{equation*}
    \inf_{\delta \in \rd(\mu_1,\phi_2)}
    \sum_{( t_1', t_2)\in S_1'\times S_2} \mu_1(t_1')\delta(t_1',t_2)
    d(t_1',t_2)\le \lambda^{ -1} \epsilon\,,
  \end{equation*}
  so there is $\delta \in \rd(\mu_1,\phi_2)$ with
  \begin{equation*}
    \sum_{(
      t_1', t_2)\in S_1'\times S_2} \mu_1(t_1')\delta(t_1',t_2)
    d(t_1',t_2)\le \lambda^{ -1} \epsilon+ \lambda^{ -1} \eta.
  \end{equation*}
  Let again $\mu_2( s)= \sum_{s_1' \in S_1} \mu_1(s_1') \delta(s_1',s)$,
  and set $L_2'( s_2, a, \mu_2)= \top$ in $P_2$.

  It is easy to see that $P_2\in \sem{ N_2}$: by construction of $P_2$,
  the identity relation $\{( s_2, s_2)\mid s_2\in S_2\}$ provides a
  refinement $P_2\preceq N_2$.  To show that $d( P_1, P_2)\le d( N_1,
  N_2)+ \eta$, we define a function $d': S_1'\times S_2\to[ 0, 1]$ by
  $d'( s_1', s_2)= d( s_1', s_2)+ \eta$ and show that $d'$ is a
  pre-fixed point to~\eqref{eq:lfpdist}.  Indeed, for $s_1'$ and $s_2$
  compatible, we have
  \begin{align*}
    d'( s_1', s_2) &= d( s_1', s_2)+ \eta \\ 
    &= \max
    \begin{cases}
      \displaystyle \max_{ a, \mu_1: L_1'( s_1', a, \mu_1)= \top\,} \min_{
        \phi_2: L_2( s_2, a, \phi_2)\ne \bot} \lambda D_{ P_1, N_2}(
      \mu_1, \phi_2, d)+ \eta \\
      \displaystyle \max_{ a, \phi_2: L_2( s_2, a, \phi_2)= \top\,} \min_{
        \mu_1: L_1'( s_1', a, \mu_1)= \top} \lambda D_{ P_1, N_2}(
      \mu_1, \phi_2, d)+ \eta
    \end{cases} \\
    &= \max
    \begin{cases}
      \displaystyle \max_{ a, \mu_1: L_1'( s_1', a, \mu_1)= \top\,} \min_{
        \mu_2: L_2'( s_2, a, \mu_2)= \top} \lambda D_{ P_1, P_2}(
      \mu_1, \mu_2, d)+ \eta \\
      \displaystyle \max_{ a, \mu_2: L_2'( s_2, a, \mu_2)= \top\,} \min_{
        \mu_1: L_1'( s_1', a, \mu_1)= \top} \lambda D_{ P_1, P_2}(
      \mu_1, \mu_2, d)+ \eta\,,
    \end{cases} \\
    \intertext{due to the construction of $P_2$ and the fact that the
      $\sup_{ \mu_1\in Sat( \mu_1)}$ is trivial in the formula for $D_{
        P_1, N_2}( \mu_1, \phi_2, d)$,}
    &\ge \max
    \begin{cases}
      \displaystyle \max_{ a, \mu_1: L_1'( s_1', a, \mu_1)= \top\,} \min_{
        \mu_2: L_2'( s_2, a, \mu_2)= \top} \lambda D_{ P_1, P_2}(
      \mu_1, \mu_2, d') \\
      \displaystyle \max_{ a, \mu_2: L_2'( s_2, a, \mu_2)= \top\,} \min_{
        \mu_1: L_1'( s_1', a, \mu_1)= \top} \lambda D_{ P_1, P_2}(
      \mu_1, \mu_2, d')\,,
    \end{cases}
  \end{align*}
  where the last inequality is a consequence of 
  \begin{equation*}
    \begin{aligned}
      \lambda D_{ P_1, P_2}( \mu_1, \mu_2, d') & = \lambda \sum_{ t_1',
        t_2} \mu_1( t_1') \delta(
      t_1', t_2)( d( t_1', t_2)+ \eta)\\
      & = \lambda \sum_{ t_1', t_2} \mu_1( t_1') \delta( t_1', t_2) d(
      t_1', t_2)+ \lambda \eta.\rlap{\hbox to 87 pt{\hfill\qEd}}
    \end{aligned}
  \end{equation*}

\section{Difference Operators for Deterministic APAs}
\label{sec:counter-ex-def}
\label{sec:diff}

The difference $N_1\setminus N_2$ of two APAs $N_1$, $N_2$ is meant to
be a syntactic representation of \emph{all counterexamples}, i.e.~all
PAs $P$ for which $P\in \impl{N_1}$ but $P\notin \impl{N_2}$.

\begin{figure}
  \subfloat[APA $N_1$]{\label{fig:c-ex-diff1}
    {\scalebox{.58}{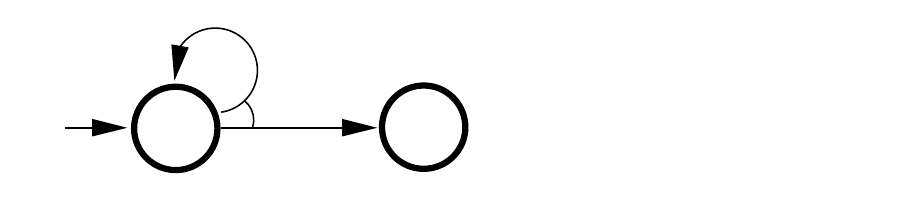}
    }}
  \hspace{.3cm}
  \subfloat[APA $N_2$]{\label{fig:c-ex-diff2}
    {\scalebox{.58}{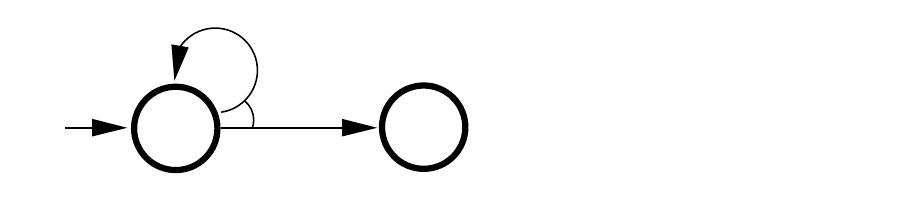}
    }}
  \caption{APAs $N_1$ and $N_2$ such that $\impl{N_1} \setminus
    \impl{N_2}$ cannot be represented using a finite-state
    APA.\label{fig:c-ex-diff}}
\end{figure}

We first observe that such a set may not be representable by an
APA. Consider the APAs $N_1$ and $N_2$ given in
Figures~\ref{fig:c-ex-diff1} and \ref{fig:c-ex-diff2}, where $\alpha \ne
\beta \ne \gamma$. Note that both $N_1$ and $N_2$ are deterministic and
in SVNF. Consider the difference of their sets of implementations. It is
easy to see that this set contains all PAs that can finitely loop on
valuation $\alpha$ and then move into a state with valuation
$\beta$. Since there is no bound on the number of steps spent in the
loop, there is no finite-state APA that can represent this set of
implementations.

By the above example, there is no hope of finding a general construction
that permits to represent the exact difference of two APAs as an APA. In
the rest of this section, we thus propose to \emph{approximate} it using
APAs. We first introduce some notations and then propose constructions
for over-approximating and under-approximating the exact difference.

\subsection{Notation}

Let $N_i = (S_i, A, L_i, AP, V_i, \{s_0^i\})$, $i = 1,2$, be
deterministic APAs in SVNF. Because $N_1$ and $N_2$ are deterministic,
we know that the difference $\impl{ N_1}\setminus \impl{ N_2}$ is
non-empty if and only if $N_1\not\preceq N_2$.  So let us assume that
$N_1\not\preceq N_2$, and let $\rel$ be a maximal refinement relation
between $N_1$ and $N_2$.  Since $N_1\not \preceq N_2$, we know that
$(s_0^1,s_0^2)\not\in\rel$. Given $(s_1,s_2) \in S_1 \times S_2$, we can
distinguish between the following cases:
\begin{enumerate}
\item $(s_1,s_2)\in \rel$,
\item $V_1(s_1)\ne V_2(s_2)$, or
\item $(s_1,s_2)\not \in \rel$ and  $V_1(s_1)= V_2(s_2)$, and 
  \begin{enumerate}
  \item 
    \begin{minipage}[t][1.5cm]{.77\linewidth}
      there exists $e\in A$ and $\phi_1\in C(S_1)$ such that
      $L_1(s_1,e,\phi_1) = \top$ and $\forall \phi_2\in
      C(S_2):L_2(s_2,e,\phi_2)=\bot$,
    \end{minipage}
    \hspace{0.2cm}
    \begin{minipage}[t][1.5cm]{0.2\linewidth}
        \raisebox{-1.4cm}{{\scalebox{0.6}{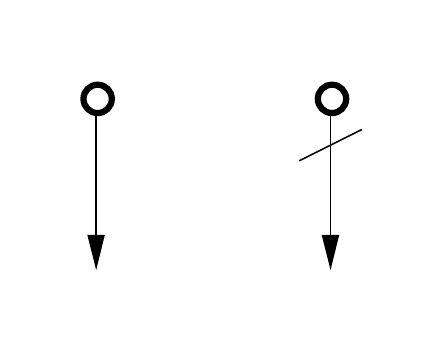}}}
      \end{minipage}

  \item 
    \begin{minipage}[t][1.5cm]{.77\linewidth}
      there exists $e\in A$ and $\phi_1\in C(S_1)$ such that
      $L_1(s_1,e,\phi_1) = \may$ and $\forall \phi_2\in
      C(S_2):L_2(s_2,e,\phi_2)=\bot$,
    \end{minipage}
    \hspace{0.2cm}
    \begin{minipage}[t][1.5cm]{0.2\linewidth}
      \raisebox{-1.4cm}{{\scalebox{0.6}{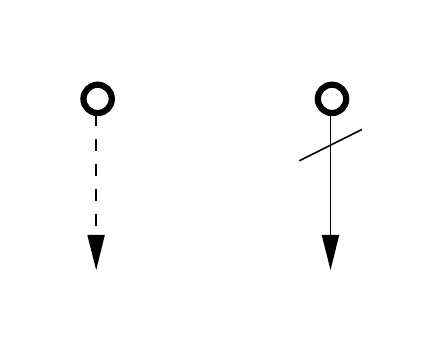}}}
    \end{minipage}

  \item 
    \begin{minipage}[t][1.5cm]{.77\linewidth}
      there exists $e\in A$ and $\phi_1\in C(S_1)$ such that
      $L_1(s_1,e,\phi_1) \ge \may$ and  $\exists \phi_2\in
      C(S_2):L_2(s_2,e,\phi_2)=\may,\exists \mu\in
      Sat(\phi_1)$ such that $\forall \mu'\in Sat(\phi_2):\mu
      \not\leqbox_{\rel} \mu'$,
    \end{minipage}
    \hspace{0.2cm}
    \begin{minipage}[t][1.5cm]{0.2\linewidth}
      \raisebox{-1.4cm}{{\scalebox{0.6}{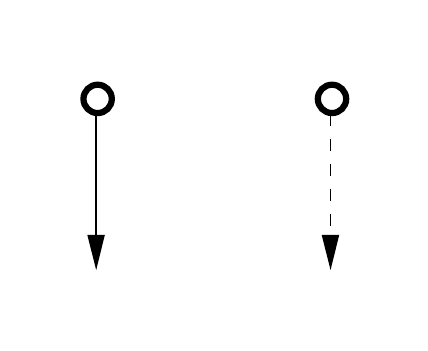}}}
    \end{minipage}

  \item 
    \begin{minipage}[t][1.5cm]{.77\linewidth}
      there exists $e\in A$ and $\phi_2\in C(S_2)$ such that
      $L_2(s_2,e,\phi_2) = \top$ and $\forall \phi_1\in
      C(S_1):L_1(s_1,e,\phi_1)=\bot$,
    \end{minipage}
    \hspace{0.2cm}
    \begin{minipage}[t][1.5cm]{0.2\linewidth}
      \raisebox{-1.4cm}{{\scalebox{0.6}{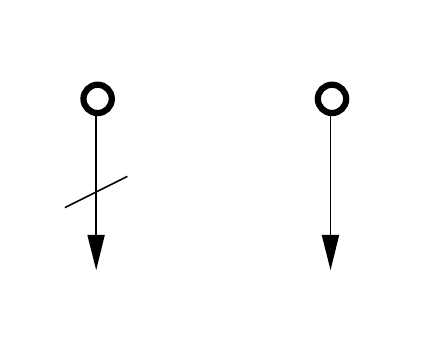}}}
    \end{minipage}

  \item 
    \begin{minipage}[t][1.5cm]{.77\linewidth}
      there exists $e\in A$ and $\phi_2\in C(S_2)$ such that
      $L_2(s_2,e,\phi_2) = \top$ and $\exists \phi_1\in
      C(S_1):L_1(s_1,e,\phi_1) = \may$,
    \end{minipage}
    \hspace{0.2cm}
    \begin{minipage}[t][1.5cm]{0.2\linewidth}
      \raisebox{-1.4cm}{{\scalebox{0.6}{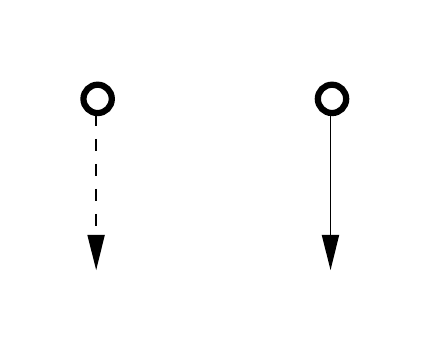}}}
    \end{minipage}

  \item 
    \begin{minipage}[t][2cm]{.77\linewidth}
      there exists $e\in A$ and $\phi_2\in C(S_2)$ such that
      $L_2(s_2,e,\phi_2) = \top$, $\exists \phi_1\in
      C(S_1):L_1(s_1,e,\phi_1) = \top$ and $ \exists \mu\in
      Sat(\phi_1)$ such that $\forall \mu'\in Sat(\phi_2):\mu
      \not\leqbox_{\rel} \mu'$.
    \end{minipage}
    \hspace{0.2cm}
    \begin{minipage}[t][2cm]{0.2\linewidth}
      \raisebox{-1.4cm}{{\scalebox{0.6}{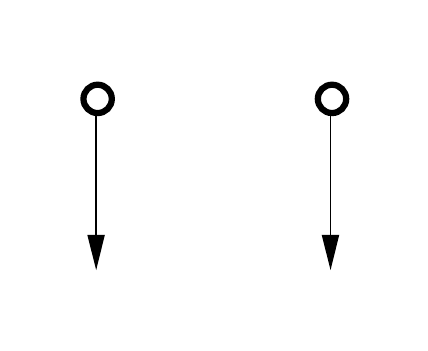}}}
    \end{minipage}

  \end{enumerate}
\end{enumerate}

\noindent Observe that because of the determinism and SVNF of APAs $N_1$ and $N_2$,
cases $1$, $2$ and $3$ cannot happen at the same time. Moreover,
although the cases in $3$ can happen simultaneously, they cannot be
``triggered'' by the same action. In order to keep track of these
``concurrent'' situations, we define the following sets.

Given a pair of states $(s_1,s_2)$, let $B_a(s_1,s_2)$ be the set of
actions in $A$ such that case $3.a$ above holds. If there is no such
action, then $B_a(s_1,s_2) = \emptyset$. Similarly, we define
$B_b(s_1,s_2), B_c(s_1,s_2), B_d(s_1,s_2), B_e(s_1,s_2)$ and
$B_f(s_1,s_2)$ to be the sets of actions such that case $3.b,c,d,e$ and
$3.f$ holds, respectively. Given a set $X \subseteq \{a,b,c,d,e,f\}$,
let $B_X(s_1,s_2) = \cup_{x \in X} B_x(s_1,s_2)$. In addition, let
$B(s_1,s_2) = B_{\{a,b,c,d,e,f\}}(s_1,s_2)$.

\subsection{Over-Approximating Difference}
\label{sec:diff-over}
\label{sec:over-approx}

We now propose a construction $\setminus^{*}$ that over-approximates the
difference between deterministic APAs in SVNF in the following sense:
given two such APAs $N_1 = (S_1, A, L_1, AP, V_1, \{s_0^1\})$ and $N_2 =
(S_2, A, L_2, AP, V_2, \{s_0^2\})$ such that $N_1 \not \preceq N_2$, we
have $\impl{N_1} \setminus \impl{N_2} \subseteq \impl{N_1 \setminus^{*}
  N_2}$.  We first observe that if $V_1(s_0^1) \ne V_2(s_0^2)$,
i.e.~$(s_0^1, s_0^2)$ in case $2$, then $\impl{N_1} \cap \impl{N_2} =
\emptyset$. In such case, we define $N_1 \setminus^{*} N_2$ as
$N_1$. Otherwise, we build on the reasons for which refinement fails
between $N_1$ and $N_2$. Note that the assumption that $N_1 \not\preceq
N_2$ implies that the pair $(s_0^1,s_0^2)$ can never be in any
refinement relation, hence in case 1. We first give an informal
intuition of how the construction works and then define it formally.

In our construction, states in $N_1 \setminus^{*} N_2$ will be elements
of $S_1 \times (S_2 \cup \{\bot\}) \times (A \cup \{\epsilon\})$. Our
objective is to ensure that any implementation of our constructed APA
will satisfy $N_1$ and not $N_2$. In $(s_1,s_2,e)$, states $s_1$ and
$s_2$ keep track of executions of $N_1$ and $N_2$. Action $e$ is the
action of $N_1$ that will be used to break satisfaction with respect to
$N_2$, i.e.~the action that will be the cause for which any
implementation of $(s_1,s_2,e)$ cannot satisfy $N_2$. Since satisfaction
is defined recursively, the breaking is not necessarily immediate and
can be postponed to successors. $\bot$ is used to represent states that
can only be reached after breaking the satisfaction relation to
$N_2$. In these states, we do not need to keep track of the
corresponding execution in $N_2$, thus only focus on satisfying
$N_1$. States of the form $(s_1,s_2,\epsilon)$ with $s_2 \ne \bot$ are
states where the satisfaction is broken by a distribution that does not
match constraints in $N_2$ (cases 3.c and 3.f). In order to invalidate
these constraints, we still need to keep track of the corresponding
execution in $N_2$, hence the use of $\epsilon$ instead of $\bot$.

The transitions in our construction will match the different cases
shown in the previous section, ensuring that in each state, either
the relation is broken immediately or reported to at least one
successor. Since there can be several ways of breaking the relation in
state $(s_0^1,s_0^2)$, each corresponding to an action $e \in
B(s_0^1,s_0^2)$, the APA $N_1 \setminus^{*} N_2$ will have one initial
state for each of them.
Formally, if $(s_0^1,s_0^2)$ is in case $3$, we define the
over-approximation of the difference of $N_1$ and $N_2$ as follows.

\begin{table}
  \caption{Definition of the transition function $L$ in
    $N_1 \setminus^{*} N_2$.\label{tab:diff}}
  \begin{tabular}{|l|c|c|l|}
    \hline
    $e \in$ & $N_1, N_2$ & $N_1 \setminus^{*} N_2$ & Formal Definition of $L$\\ \hline

    $B_a(s_1,s_2)$  & 
    \begin{minipage}[c][1.5cm]{0.13\linewidth}
      \raisebox{-1.4cm}{\hspace{-0.2cm}{\scalebox{0.4}{\input{case3a.pdf_t}}}}
    \end{minipage}
    &
    \multirow{2}{*}{\begin{minipage}[c][2cm]{0.12\linewidth}
      \raisebox{-1.6cm}{{\scalebox{0.47}{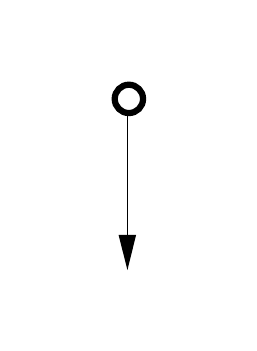}}}
    \end{minipage}}
    & 
    \multirow{2}{*}{\begin{minipage}[c][2.3cm]{0.6\linewidth}         
        For all $a \ne e \in A$ and $\phi \in C(S_1)$ such that
        $L_1(s_1, a ,\phi) \ne \bot$, let
        $L((s_1,s_2,e),a,\phi^{\bot}) = L_1(s_1, a ,\phi)$. In
        addition, let $L((s_1,s_2,e),e,\phi_1^{\bot}) = \top$. For all
        other $b \in A$ and $\phi \in C(S)$, let
        $L((s_1,s_2,e),b,\phi) = \bot$.
    \end{minipage}}

\\\cline{1-2}

    $B_b(s_1,s_2)$  & 
    \begin{minipage}[c][1.5cm]{0.13\linewidth}
      \raisebox{-1.4cm}{\hspace{-0.2cm}{\scalebox{0.4}{\input{case3b.pdf_t}}}}
    \end{minipage}
    &
    &

\\\hline

    $B_d(s_1,s_2)$  & 
    \begin{minipage}[c][1.8cm]{0.13\linewidth}
      \raisebox{-1.4cm}{\hspace{-0.2cm}{\scalebox{0.4}{\input{case3d.pdf_t}}}}
    \end{minipage}
    &
    \begin{minipage}[c][1.8cm]{0.12\linewidth}
      \raisebox{-1.4cm}{{\scalebox{0.47}{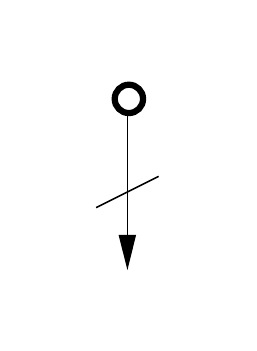}}}
    \end{minipage}
    &
    \begin{minipage}[c][1.8cm]{0.6\linewidth}
     For all $a \in A$ and $\phi \in C(S_1)$ such that $L_1(s_1, a
     ,\phi) \ne \bot$, let $L((s_1,s_2,e),a,\phi^{\bot}) =
     L_1(s_1, a ,\phi)$. For all other $b \in A$ and $\phi \in
     C(S)$, let $L((s_1,s_2,e),b,\phi) = \bot$.
    \end{minipage}

\\\hline

    $B_e(s_1,s_2)$  & 
    \begin{minipage}[c][2.3cm]{0.13\linewidth}
      \raisebox{-1.4cm}{\hspace{-0.2cm}{\scalebox{0.4}{\input{case3e.pdf_t}}}}
    \end{minipage}
    &
    \begin{minipage}[c][2.3cm]{0.12\linewidth}
      \raisebox{-1.4cm}{{\scalebox{0.47}{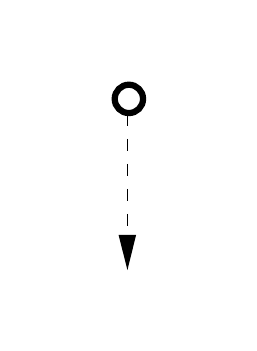}}}
    \end{minipage}
    &
    \begin{minipage}[c][2.3cm]{0.6\linewidth}
      For all $a \ne e \in A$ and $\phi \in C(S_1)$ such that
      $L_1(s_1, a ,\phi) \ne \bot$, let
      $L((s_1,s_2,e),a,\phi^{\bot}) = L_1(s_1, a ,\phi)$.
      In addition, let $L((s_1,s_2,e),e,\phi^B_{12}) = \may$.
      For all other $b \in A$ and $\phi \in C(S)$, let
      $L((s_1,s_2,e),b,\phi) = \bot$.
    \end{minipage}

\\\hline

    $B_c(s_1,s_2)$  & 
    \begin{minipage}[c][1.5cm]{0.13\linewidth}
      \raisebox{-1.4cm}{\hspace{-0.2cm}{\scalebox{0.4}{\input{case3c.pdf_t}}}}
    \end{minipage}
    &
    \multirow{2}{*}{\begin{minipage}[c][2cm]{0.12\linewidth}
      \raisebox{-1.4cm}{{\scalebox{0.47}{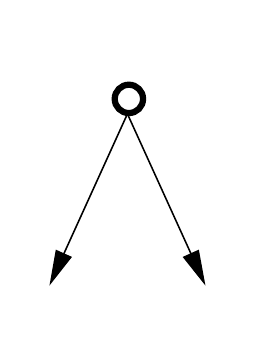}}}
    \end{minipage}}
    &
    \multirow{2}{*}{\begin{minipage}[c][2.3cm]{0.6\linewidth}
      For all $a \in A$ and $\phi \in C(S_1)$ such that $L_1(s_1, a
      ,\phi) \ne \bot$ (including $e$ and $\phi_1$), let
      $L((s_1,s_2,e),a,\phi^{\bot}) = L_1(s_1, a ,\phi)$.
      In addition, let $L((s_1,s_2,e),e,\phi^B_{12}) = \top$.
      For all other $b \in A$ and $\phi \in C(S)$, let
      $L((s_1,s_2,e),b,\phi) = \bot$.
    \end{minipage}}

\\\cline{1-2}

    $B_f(s_1,s_2)$  & 
    \begin{minipage}[c][1.5cm]{0.13\linewidth}
      \raisebox{-1.4cm}{\hspace{-0.2cm}{\scalebox{0.4}{\input{case3f.pdf_t}}}}
    \end{minipage}
    &
    &
    \\\hline

  \end{tabular}
\end{table}

\begin{defi}
  Let $N_1 \setminus^{*} N_2 = (S,A,L,AP,V,S_0)$, where $S = S_1 \times
  (S_2 \cup \{\bot\}) \times (A \cup \{\epsilon\})$, $V(s_1,s_2,a) =
  V(s_1)$ for all $s_2$ and $a$, $S_0 = \{(s_0^1,s_0^2, f) \st f \in
  B(s_0^1, s_0^2)\}$, and $L$ is defined by:
\begin{itemize}
\item If $s_2 = \bot$ or $e = \epsilon$ or $(s_1, s_2)$ in case $1$ or
  $2$, then for all $a \in A$ and $\phi \in C(S_1)$ such that
  $L_1(s_1, a ,\phi) \ne \bot$, let $L((s_1,s_2,e),a,\phi^{\bot}) =
  L_1(s_1, a ,\phi)$, with $\phi^\bot$ defined below. For all other $b
  \in A$ and $\phi \in C(S)$, let $L((s_1,s_2,e),b,\phi) = \bot$.
\item Else, we have $(s_1,s_2)$ in case $3$ and $B(s_1,s_2) \ne
  \emptyset$ by construction. The definition of $L$ is given in
  Table~\ref{tab:diff}, with the constraints $\phi^\bot$ and
  $\phi^B_{12}$ defined hereafter.
\end{itemize}
\end{defi}\medskip

\noindent Given $\phi \in C(S_1)$, $\phi^{\bot} \in C(S)$ is defined as follows:
$\mu \in Sat(\phi^{\bot})$ iff $\forall s_1 \in S_1, \forall s_2 \ne
\bot, \forall b \ne \epsilon, \mu(s_1,s_2,b) = 0$ and the distribution
$(\mu\downarrow_1 : s_1 \mapsto \mu(s_1,\bot,\epsilon))$ is in
$Sat(\phi)$.  

Given a state $(s_1,s_2,e) \in S$ with $s_2 \ne \bot$ and $e \ne
\epsilon$ and two constraints $\phi_1 \in C(S_1)$, $\phi_2 \in C(S_2)$
such that $L_1(s_1,e,\phi_1) \ne \bot$ and $L_2(s_2,e,\phi_2) \ne \bot$,
the constraint $\phi^B_{12} \in C(S)$ is defined as follows: $\mu \in
Sat(\phi^B_{12})$ iff
\begin{enumerate}
\item for all $(s'_1,s'_2,c) \in S$, we have $\mu(s'_1, s'_2,c) >0
  \Rightarrow s'_2 = \bot$ if $\textsf{succ}_{s_2,e}(s_1') = \emptyset$
  and $s'_2 = \textsf{succ}_{s_2,e}(s_1')$ otherwise, and $c \in
  B(s'_1,s'_2) \cup \{\epsilon\}$,
\item the distribution $\mu_1 : s'_1 \mapsto \sum_{c \in
    A\cup\{\epsilon\}, s'_2 \in S_2\cup \{\bot\}} \mu(s'_1,s'_2,c)$
  satisfies $\phi_1$, and
\item one of the following holds:
  \begin{enumerate}
  \item there exists $(s'_1, \bot, c)$ such that $\mu(s'_1, \bot, c) >0$,
  \item the distribution $\mu_2 : s'_2 \mapsto \sum_{c \in
      A\cup\{\epsilon\}, s'_1 \in S_1} \mu(s'_1,s'_2,c)$ does not
    satisfy $\phi_2$, or
  \item there exists $s'_1 \in S_1$, $s'_2 \in S_2$ and $c \ne \epsilon$
    such that $\mu(s'_1,s'_2,c)>0$.
  \end{enumerate}
\end{enumerate}
Informally, distributions in $\phi^B_{12}$ must (1) follow the
corresponding execution is $N_1$ and $N_2$ if possible, (2) satisfy
$\phi_1$ and (3), (a) reach a state in $N_1$ that cannot be matched in
$N_2$, (b) break the constraint $\phi_2$, or (c) report breaking the
relation to at least one successor state.

The following theorem shows that $N_1 \setminus^{*} N_2$ is, as
intended, an over-approximation of the difference of $N_1$ and $N_2$ in
terms of sets of implementations.

\begin{thm}
  \label{th:over-diff}
  For all deterministic APAs $N_1$ and $N_2$ in SVNF such that $N_1 \not
  \preceq N_2$, we have $\impl{N_1} \setminus \impl{N_2} \subseteq
  \impl{N_1 \setminus^{*} N_2}$.
\end{thm}

\proof
  Let $N_1=(S_1,A,L_1,AP,V_1,\{s_0^1\})$ and
  $N_2=(S_2,A,L_2,AP,V_2,\{s_0^2\})$ be deterministic APAs in SVNF such
  that $N_1\not\preceq N_2$.  Let $\rel$ be the maximal refinement
  relation between $N_1$ and $N_2$. Let $P = (S_P, A, L_P, AP, V_P,
  s_0^P)$ be a PA such that $P \sat N_1$ and $P \not \sat N_2$. We prove
  that $P \sat N_1 \setminus^{*} N_2$. Let $\rel_1 \subseteq S_P \times
  S_1$ be the relation witnessing $P \sat N_1$ and let $\rel_2$ be the
  maximal satisfaction relation in $S_P \times S_2$. By construction,
  $(s_0^P, s_2) \notin \rel_2$.

  If $V_1(s_0^1) \ne V_2(s_0^2)$, then by construction $N_1
  \setminus^{*} N_2 = N_1$ and thus $P \sat N_1 \setminus^{*} N_2$.
  Else, we have $(s_0^1,s_0^2)$ in case 3, thus $N_1 \setminus^{*} N_2 =
  (S,A,L,AP,V,S_0)$ is defined as in Section~\ref{sec:diff-over}. By
  construction, we also have $(s_0^P,s_0^2)$ in case 3, thus there must
  exist $f \in B(s_0^P,s_0^2)$. Observe that by construction, we must
  have $B(s_0^P, s_0^2) \subseteq B(s_0^1,s_0^2)$. We will prove that $P
  \sat N_1 \setminus^{*} N_2$.
  Define the following relation $\rel^{\setminus} \subseteq S_P \times
  S$:
  \begin{equation*}
    p \rel^{\setminus} (s_1,s_2,e) \iff \left \{ 
      \begin{array}{cl}
        &  (p \rel_1 s_1) \text{ and } (s_2 = \bot) \text{ and } (e =
        \varepsilon) \\
        \text{or} & (p \rel_1 s_1) \text{ and } (p,s_2) \text{ in case 1
          or 2 and } \text{ and } (e = \varepsilon) \\
        \text{or} & (p \rel_1 s_1) \text{ and } (p,s_2) \text{ in case 3
          and } (e \in B(p, s_2))
      \end{array}
    \right.
  \end{equation*}

  \noindent We now prove that $\rel^{\setminus}$ is a satisfaction relation. Let
  $(p,(s_1,s_2,e)) \in \rel^{\setminus}$.

  If $s_2 = \bot$ or $e = \varepsilon$, then since $p \rel_1 s_1$,
  $\rel^\setminus$ satisfies the axioms of a satisfaction relation by
  construction.  Else we have $s_2 \in S_2$ and $e \ne \epsilon$, thus,
  by definition of $\rel^{\setminus}$, we know that $(p,s_2)$ is in case
  $3$.
  \begin{itemize}
  \item By construction, we have $V_P(p) \in V_1(s_1) = V((s_1,s_2,e))$.

  \item Let $a \in A$ and $\mu_P \in Dist(S_P)$ such that
    $L_P(p,a,\mu_P) = \top$. There are several cases.
    \begin{itemize}
    \item If $a \ne e$, then since $p \rel_1 s_1$, there exists $\phi_1
      \in C(S_1)$ such that $L_1(s_1,a,\phi_1) \ne \bot$ and there
      exists $\mu_1 \in Sat(\phi_1)$ such that $\mu_P
      \leqbox_{\rel^{\setminus}} \mu_1$. By construction, we have
      $L((s_1,s_2,e),a,\phi_1^{\bot}) \ne \bot$ and there obviously
      exists $\mu \in Sat(\phi_1^{\bot})$ such that $\mu_P
      \leqbox_{\rel^{\setminus}} \mu$.

    \item If $a = e \in B_a(p,s_2)$, then, as above, there exists a
      constraint $\phi \in C(S)$ such that $L((s_1,s_2,e),a,\phi) \ne
      \bot$ and there exists $\mu \in Sat(\phi)$ such that $\mu_P
      \leqbox_{\rel^{\setminus}} \mu$. Observe that $B_a(s_1,s_2)
      \subseteq B_a(p,s_2) \subseteq B_a(s_1,s_2)\cup B_b(s_1,s_2)$.

    \item Else, we necessarily have $a = e \in B_c(p,s_2) \cup
      B_f(p,s_2)$. Observe that, by construction, $B_c(p,s_2) \subseteq
      B_c(s_1,s_2)$ and $B_f(p,s_2) \subseteq B_f(s_1,s_2)$. Since $p
      \rel_1 s_1$, there exists $\phi_1 \in C(S_1)$ such that
      $L_1(s_1,e,\phi_1) \ne \bot$ and there exists $\mu_1 \in
      Sat(\phi_1)$ and a correspondence function $\delta_1 : S_P
      \rightarrow (S_1 \rightarrow [0,1])$ such that $\mu_P
      \leqbox_{\rel_1}^{\delta_1} \mu_1$.

      Moreover, by construction of $N_1 \setminus^{*} N_2$, we know that
      the constraint $\phi_{12}^B$ such that $\mu \in Sat(\phi_{12}^B)$
      iff. (1) for all $(s'_1,s'_2,c) \in S$, we have $\mu(s'_1, s'_2,c)
      >0 \Rightarrow s'_2 = \bot$ if $\textsf{succ}_{s_2,e}(s_1') =
      \emptyset$ and $s'_2 = \textsf{succ}_{s_2,e}(s_1')$ otherwise, and
      $c \in B(s'_1,s'_2) \cup \{\epsilon\}$, (2) the distribution
      $\mu_1 : s'_1 \mapsto \sum_{c \in A\cup\{\epsilon\}, s'_2 \in
        S_2\cup \{\bot\}} \mu(s'_1,s'_2,c)$ satisfies $\phi_1$, and (3)
      either (b) the distribution $\mu_2 : s'_2 \mapsto \sum_{c \in
        A\cup\{\epsilon\}, s'_1 \in S_1} \mu(s'_1,s'_2,c)$ does not
      satisfy $\phi_2$, or (c) there exists $s'_1 \in S_1$, $s'_2 \in
      S_2$ and $c \ne \epsilon$ such that $\mu(s'_1,s'_2,c)>0$ is such
      that $L((s_1,s_2,e),e,\phi_{12}^B) = \top$.

      We now prove that there exists $\mu \in Sat(\phi_{12}^B)$ such
      that $\mu_P \leqbox_{\rel^{\setminus}} \mu$. Consider the function
      $\delta^{\setminus} : S_P \rightarrow (S \rightarrow [0,1])$
      defined as follows: Let $p' \in S_P$ such that $\mu_P(p') >0$ and
      let $s'_1 = \textsf{succ}_{s_1,e}(p')$, which exists by $\rel_1$.
      \begin{itemize}
      \item If $\textsf{succ}_{s_2,e}(p') = \emptyset$, then
        $\delta^{\setminus}(p')(s'_1,\bot,\epsilon) = 1$.
      \item Else, let $s'_2 = \textsf{succ}_{s_2,e}(p')$. Then,
        \begin{itemize}
        \item if $(p',s'_2) \in \rel_2$, then
          $\delta^{\setminus}(p')(s'_1,s'_2,\epsilon) = 1$.
        \item Else, $(p',s'_2)$ is in case 3 and $B(p',s'_2) \ne
          \emptyset$. In this case, let $c \in B(p',s'_2)$ and define
          $\delta^{\setminus}(p',(s'_1,s'_2,c)) = 1$. For all other $c'
          \in B(p',s'_2)$, define $\delta^{\setminus}(p',(s'_1,s'_2,c))
          = 0$.
        \end{itemize}
      \end{itemize}

      Observe that for all $p' \in S_P$ such that $\mu_P(p') >0$, there
      exists a unique $s' \in S'$ such that $\delta^{\setminus}(p')(s')
      = 1$. Thus $\delta^{\setminus}$ is a correspondence function.

      We now prove that $\mu = \mu_P \delta^{\setminus} \in
      Sat(\phi_{12}^B)$.
      \begin{enumerate}
      \item Let $(s'_1,s'_2,c) \in S$ such that $\mu(s'_1, s'_2,c)
        >0$. By construction, there exists $p' \in S_P$ such that
        $\mu_P(p') >0$ and $\delta^{\setminus}(p')(s'_1,s'_2,c)
        >0$. Moreover, $c \in B(s'_1,s'_2) \cup \{\epsilon\}$, and $s'_2
        = \bot$ if $\textsf{succ}_{s_2,e}(s_1') = \emptyset$ and $s'_2 =
        \textsf{succ}_{s_2,e}(s_1')$ otherwise.

      \item Consider the distribution $\mu'_1:s'_1 \mapsto \sum_{c \in
          A\cup\{\epsilon\}, s'_2 \in S_2\cup \{\bot\}}
        \mu(s'_1,s'_2,c)$. By determinism (See Lemma 28
        in~\cite{TCS11}), we have that $\delta_1(p')(s'_1) = 1 \iff s'_1
        = \textsf(succ)_{s_1,e}(p')$. As a consequence, we have that
        $\mu'_1 = \mu_1 \in Sat(\phi_1)$.

      \item Assume that for all $p' \in S_P$ such that $\mu_P(p') >0$,
        we have $\textsf{succ}_{s_2,e}(p') \ne \emptyset$ (the other
        case being trivial). Consider the distribution $\mu_2 : s'_2
        \mapsto \sum_{c \in A\cup\{\epsilon\}, s'_1 \in S_1}
        \mu(s'_1,s'_2,c)$ and let $\delta_2 : S_P \rightarrow (S_2
        \rightarrow [0,1])$ be such that $\delta_2(p')(s'_2) = 1 \iff
        s'_2 = \textsf{succ}_{s_2,e}(p')$. By construction, $\delta_2$
        is a correspondence function and $\mu_2 = \mu_P \delta_2$.
        Since $e \in B_c(p,s_2) \cup B_f(p,s_2)$, we have that $\mu_P
        \not \leqbox_{\rel_2} \mu_2$. If $\mu_2 \notin Sat(\phi_2)$,
        then we have $\mu \in Sat(\phi_{12}^B)$. Else, there must exist
        $p' \in S_P$ and $s'_2 \in S_2$ such that $\mu_P(p') >0$,
        $\delta_2(p')(s'_2) >0$ and $(p',s'_2) \notin \rel_2$. As a
        consequence, $(p',s'_2)$ is in case $3$ and there exists $c \ne
        \epsilon$ such that $\delta^{\setminus}(p')(s'_1,s'_2,c) >0$,
        thus $\mu(s'_1,s'_2,c) >0$. As a consequence, $\mu \in
        Sat(\phi_{12}^B)$.
      \end{enumerate}

      We thus conclude that there exists $\mu \in Sat(\phi_{12}^B)$ such
      that $\mu_P \leqbox_{\rel^{\setminus}} \mu$.
    \end{itemize}

    Finally, in all cases, there exists $\phi \in C(S)$ such that
    $L((s_1,s_2,e),a,\phi) \ne \bot$ and there exists $\mu \in
    Sat(\phi)$ such that $\mu_P \leqbox_{\rel^{\setminus}} \mu$.
     
  \item Let $a \in A$ and $\phi \in C(S)$ such that
    $L((s_1,s_2,e),a,\phi) = \top$. As above, there are several cases.
    \begin{itemize}
    \item If $a \ne e$, then, by construction of $N_1 \setminus^{*}
      N_2$, there must exists $\phi_1 \in C(S_1)$ such that
      $L_1(s_1,a,\phi_1) = \top$. The rest of the proof is then as
      above.

    \item If $a = e \in B_a(p,s_2)$, then there exists $\mu_P \in
      Dist(S_P)$ such that $L_P(p,e,\mu_P) = \top$. The rest of the
      proof is then as above. Recall that $B_a(s_1,s_2) \subseteq
      B_a(p,s_2) \subseteq B_a(s_1,s_2)\cup B_b(s_1,s_2)$.

    \item Else, we necessarily have $a = e \in B_c(p,s_2) \cup
      B_f(p,s_2)$. Recall that, by construction, $B_c(p,s_2) \subseteq
      B_c(s_1,s_2)$ and $B_f(p,s_2) \subseteq B_f(s_1,s_2)$. Thus, there
      exists $\mu_P \in Dist(S_P)$ and $\phi_2 \in C(S_2)$ such that
      $L_2(s_2,e,\phi_2) \ne \bot$ and $\forall \mu_2 \in Sat(\phi_2),
      \mu_P \not \leqbox_{\rel_2} \mu_2$. Since $e \in B_c(s_1,s_2) \cup
      B_f(s_1,s_2)$, there also exist $\phi_1 \in C(S_1)$ such that
      $L_1(s_1,e,\phi_1) \ne \bot$. By determinism, $\phi_1$ and
      $\phi_2$ are unique. The rest of the proof follows as above.
    \end{itemize}

    \noindent Thus, in all cases, there exists $\mu_P \in Dist(S_P)$ such that
    $L_P(p,a,\mu_P) = \top$ and there exists $\mu \in Sat(\phi)$ such
    that $\mu_P \leqbox_{\rel^{\setminus}} \mu$.
  \end{itemize}

  \noindent Finally, $\rel^{\setminus}$ is a satisfaction relation. Moreover, we
  have $s_0^P \rel_1 s_0^1$, $(s_0^P, s_0^2)$ in case 3 and $f \in
  B(s_0^P,s_0^2)$ by construction, thus $s_0^P \rel^{\setminus}
  (s_0^1,s_0^2, f) \in S_0$.  We thus conclude that $P \sat N_1
  \setminus^{*} N_2$.
\qed

The reverse inclusion unfortunately does not hold. Intuitively, as
explained in the construction of the constraint $\phi^B_{12}$ above, one
can postpone the breaking of the satisfaction relation for $N_2$ to the
next state (condition (3.c)). This assumption is necessary in order to
produce an APA representing \emph{all} counterexamples. However, when
there are cycles in the execution of $N_1 \setminus^{*} N_2$, then we
may postpone forever, thus allowing for implementations that will
ultimately satisfy $N_2$. This is illustrated in the following example.

\begin{figure}
\centering
\subfloat[\ $N_1 \setminus^{*} N_2$]{\label{fig:ex-over-approx1}\scalebox{.55}{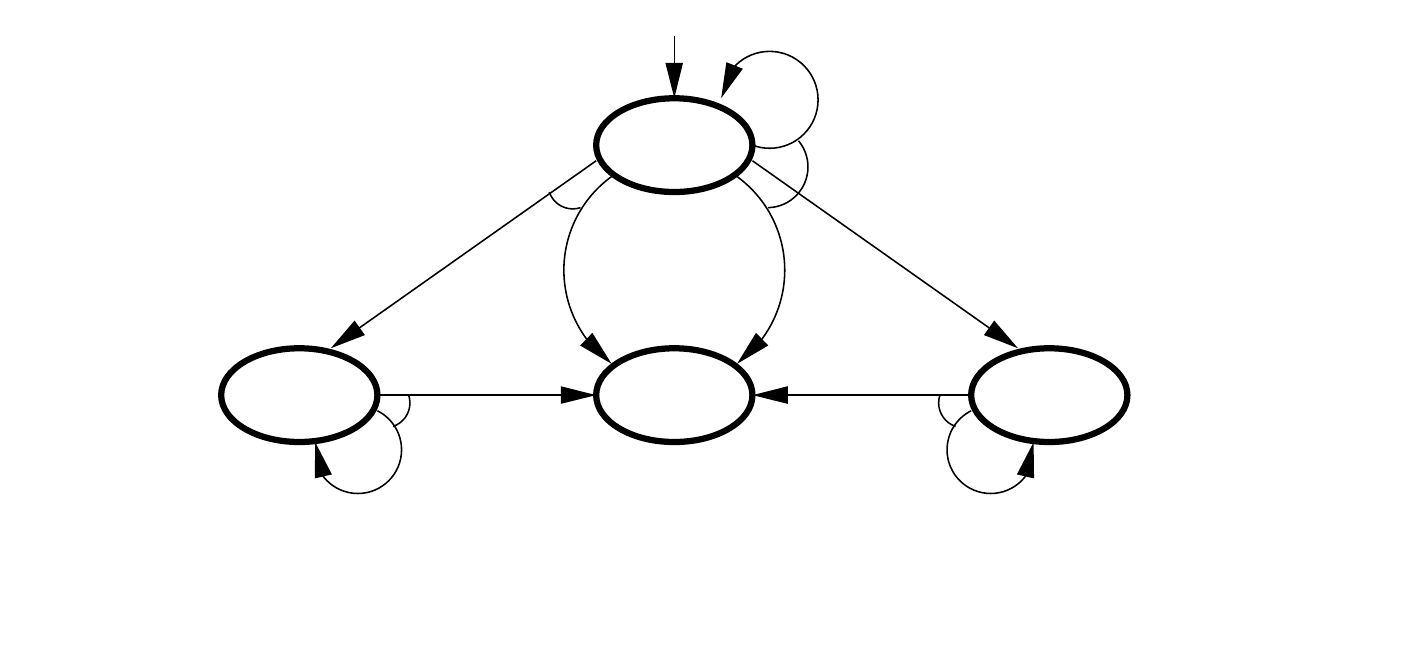}}
\hspace{1cm}
\subfloat[\ $P$]{\label{fig:ex-over-approx2}\scalebox{.55}{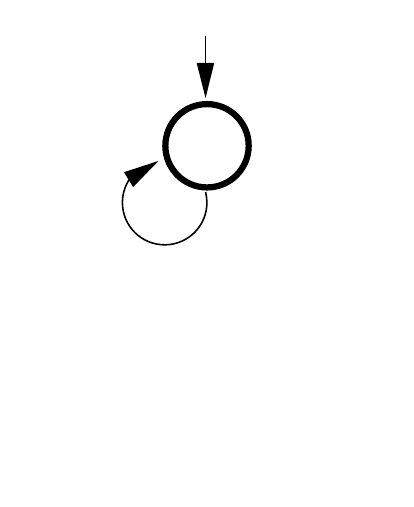}}
\caption{Over-approximating difference $N_1 \setminus^{*} N_2$ of APAs
  $N_1$ and $N_2$ from Figure~\ref{fig:c-ex-diff} and PA $P$ such that
  $P \sat N_1 \setminus^{*} N_2$ and $P \sat N_2$.}
\label{fig:ex-over-approx}
\end{figure}

\begin{exa}
  Consider the APAs $N_1$ and $N_2$ given in
  Fig.~\ref{fig:c-ex-diff}. Their over-approximating difference $N_1
  \setminus^{*} N_2$ is given in Fig.~\ref{fig:ex-over-approx1}. One can
  see that the PA $P$ in Fig.~\ref{fig:ex-over-approx2} satisfies both
  $N_1 \setminus^{*} N_2$ and $N_2$.
\end{exa}

We will later see in Corollary~\ref{co:over-approx-best} that even
though $N_1\setminus^* N_2$ may be capturing too many counterexamples,
the \emph{distance} between $N_1\setminus^* N_2$ and the real set of
counterexamples $\sem{ N_1}\setminus \sem{ N_2}$ is zero.  This means
that the two sets are infinitesimally close to each other, so in this
sense, and with respect to this distance, $N_1\setminus^* N_2$ is a
\emph{best possible} over-approximation.

\subsection{Under-Approximating Difference}
\label{sec:under-approx}

We now propose a construction that instead \emph{under-estimates} the
difference between APAs. This construction resembles the
over-approx\-imation presented in the previous section, the main
difference being that in the under-approximation, states are indexed
with integers which represent the maximal depth of the unfolding of
counterexamples.  The construction is as follows.

Let $N_1 = (S_1, A, L_1, AP, V_1, \{s_0^1\})$ and $N_2 = (S_2, A, L_2,
AP, V_2, \{s_0^2\})$ be two deterministic APAs in SVNF such that $N_1
\not \preceq N_2$. Let $K \in \mathbb{N}$ be the parameter of our
construction. As in Section~\ref{sec:over-approx}, if $V_1(s_0^1) \ne
V_2(s_0^2)$, i.e.~$(s_0^1, s_0^2)$ in case $2$, then $\impl{N_1} \cap
\impl{N_2} = \emptyset$. In this case, we define $N_1 \setminus^{K} N_2$
as $N_1$. Otherwise, the under-approximation is defined as follows.

\begin{defi}\label{def:under-approx}
  Let $N_1 \setminus^{K} N_2 = (S,A,L,AP,V,S^K_0)$, where $S = S_1
  \times (S_2 \cup \{\bot\}) \times (A \cup \{\epsilon\}) \times \{1,
  \ldots, K\}$, $V(s_1,s_2,a, k) = V(s_1)$ for all $s_2$, $a$, $k< K$,
  $S^K_0 = \{(s_0^1,s_0^2, f, K) \st f \in B(s_0^1, s_0^2)\}$, and $L$
  is defined by:
\begin{itemize}
\item If $s_2 = \bot$ or $e = \epsilon$ or $(s_1, s_2)$ in case $1$ or
  $2$, then for all $a \in A$ and $\phi \in C(S_1)$ such that
  $L_1(s_1, a ,\phi) \ne \bot$, let $L((s_1,s_2,e, k),a,\phi^{\bot}) =
  L_1(s_1, a ,\phi)$, with $\phi^\bot$ defined below. For all other $b
  \in A$ and $\phi \in C(S)$, let $L((s_1,s_2,e, k),b,\phi) = \bot$.
\item Else we have $(s_1,s_2)$ in case $3$ and $B(s_1,s_2) \ne
  \emptyset$ by construction. The definition of $L$ is given in
  Table~\ref{tab:under-diff}. The constraints $\phi^\bot$ and
  $\phi_{12}^{B,k}$ are defined hereafter.
\end{itemize}
\end{defi}

\begin{table}
  \caption{%
    \label{tab:under-diff}
    Definition of the transition function $L$ in
    $N_1 \setminus^{K} N_2$.}
  \begin{tabular}{|l|c|c|l|}
    \hline
    $e \in$ & $N_1, N_2$ & $N_1 \setminus^{K} N_2$ & Formal Definition of $L$\\ \hline

    $B_a(s_1,s_2)$  & 
    \begin{minipage}[c][1.5cm]{0.13\linewidth}
      \raisebox{-1.4cm}{\hspace{-0.2cm}{\scalebox{0.4}{\input{case3a.pdf_t}}}}
    \end{minipage}
    &
    \multirow{2}{*}{\begin{minipage}[c][2cm]{0.12\linewidth}
      \raisebox{-1.6cm}{{\scalebox{0.47}{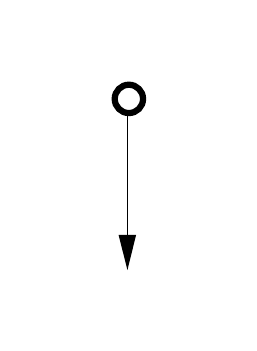}}}
    \end{minipage}}
    & 
    \multirow{2}{*}{\begin{minipage}[c][2.3cm]{0.6\linewidth}         
        For all $a \ne e \in A$ and $\phi \in C(S_1)$ such that
        $L_1(s_1, a ,\phi) \ne \bot$, let
        $L((s_1,s_2,e,k),a,\phi^{\bot}) = L_1(s_1, a ,\phi)$. In
        addition, let $L((s_1,s_2,e,k),e,\phi_1^{\bot}) = \top$. For all
        other $b \in A$ and $\phi \in C(S)$, let
        $L((s_1,s_2,e,k),b,\phi) = \bot$.
    \end{minipage}}

\\\cline{1-2}

    $B_b(s_1,s_2)$  & 
    \begin{minipage}[c][1.5cm]{0.13\linewidth}
      \raisebox{-1.4cm}{\hspace{-0.2cm}{\scalebox{0.4}{\input{case3b.pdf_t}}}}
    \end{minipage}
    &
    &

\\\hline

    $B_d(s_1,s_2)$  & 
    \begin{minipage}[c][1.8cm]{0.13\linewidth}
      \raisebox{-1.4cm}{\hspace{-0.2cm}{\scalebox{0.4}{\input{case3d.pdf_t}}}}
    \end{minipage}
    &
    \begin{minipage}[c][1.8cm]{0.12\linewidth}
      \raisebox{-1.4cm}{{\scalebox{0.47}{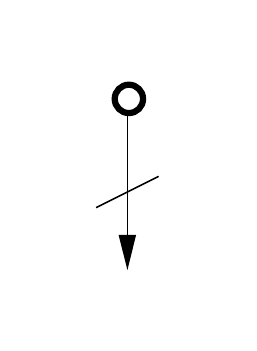}}}
    \end{minipage}
    &
    \begin{minipage}[c][1.8cm]{0.6\linewidth}
     For all $a \in A$ and $\phi \in C(S_1)$ such that $L_1(s_1, a
     ,\phi) \ne \bot$, let $L((s_1,s_2,e,k),a,\phi^{\bot}) =
     L_1(s_1, a ,\phi)$. For all other $b \in A$ and $\phi \in
     C(S)$, let $L((s_1,s_2,e,k),b,\phi) = \bot$.
    \end{minipage}

\\\hline

    $B_e(s_1,s_2)$  & 
    \begin{minipage}[c][2.3cm]{0.13\linewidth}
      \raisebox{-1.4cm}{\hspace{-0.2cm}{\scalebox{0.4}{\input{case3e.pdf_t}}}}
    \end{minipage}
    &
    \begin{minipage}[c][2.3cm]{0.12\linewidth}
      \raisebox{-1.4cm}{{\scalebox{0.47}{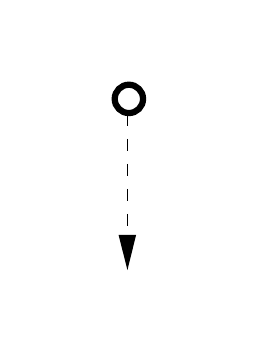}}}
    \end{minipage}
    &
    \begin{minipage}[c][2.3cm]{0.6\linewidth}
      For all $a \ne e \in A$ and $\phi \in C(S_1)$ such that
      $L_1(s_1, a ,\phi) \ne \bot$, let
      $L((s_1,s_2,e,k),a,\phi^{\bot}) = L_1(s_1, a ,\phi)$.
      In addition, let $L((s_1,s_2,e,k),e,\phi^{B,k}_{12}) = \may$.
      For all other $b \in A$ and $\phi \in C(S)$, let
      $L((s_1,s_2,e,k),b,\phi) = \bot$.
    \end{minipage}

\\\hline

    $B_c(s_1,s_2)$  & 
    \begin{minipage}[c][1.5cm]{0.13\linewidth}
      \raisebox{-1.4cm}{\hspace{-0.2cm}{\scalebox{0.4}{\input{case3c.pdf_t}}}}
    \end{minipage}
    &
    \multirow{2}{*}{\begin{minipage}[c][2cm]{0.12\linewidth}
      \raisebox{-1.4cm}{{\scalebox{0.47}{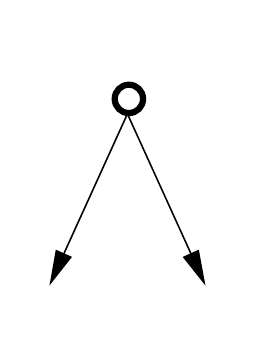}}}
    \end{minipage}}
    &
    \multirow{2}{*}{\begin{minipage}[c][2.3cm]{0.6\linewidth}
      For all $a \in A$ and $\phi \in C(S_1)$ such that $L_1(s_1, a
      ,\phi) \ne \bot$ (including $e$ and $\phi_1$), let
      $L((s_1,s_2,e,k),a,\phi^{\bot}) = L_1(s_1, a ,\phi)$.
      In addition, let $L((s_1,s_2,e,k),e,\phi^{B,k}_{12}) = \top$.
      For all other $b \in A$ and $\phi \in C(S)$, let
      $L((s_1,s_2,e,k),b,\phi) = \bot$.
    \end{minipage}}

\\\cline{1-2}

    $B_f(s_1,s_2)$  & 
    \begin{minipage}[c][1.5cm]{0.13\linewidth}
      \raisebox{-1.4cm}{\hspace{-0.2cm}{\scalebox{0.4}{\input{case3f.pdf_t}}}}
    \end{minipage}
    &
    &
    \\\hline

  \end{tabular}
\end{table}

\noindent Given a constraint $\phi \in C(S_1)$, the constraint $\phi^{\bot} \in
C(S)$ is defined as follows: $\mu \in Sat(\phi^{\bot})$ iff $\forall s_1
\in S_1, \forall s_2 \ne \bot, \forall b \ne \epsilon, \forall k \ne 1,
\mu(s_1,s_2,b,k) = 0$ and the distribution $(\mu\downarrow_1 : s_1
\mapsto \mu(s_1,\bot,\epsilon, 1))$ is in $Sat(\phi)$.

\begin{figure}
\centering
\subfloat[\quad $N_1 \setminus^1 N_2$]{\label{fig:under-approx1}\scalebox{.55}{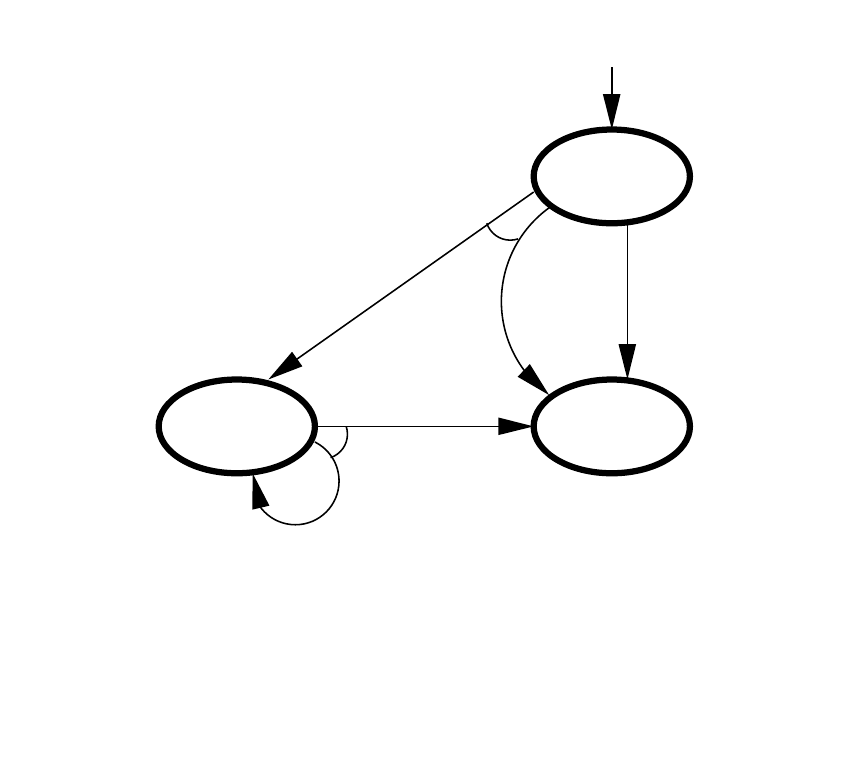}}
\subfloat[\quad $N_1 \setminus^2 N_2$]{\label{fig:under-approx2}\scalebox{.55}{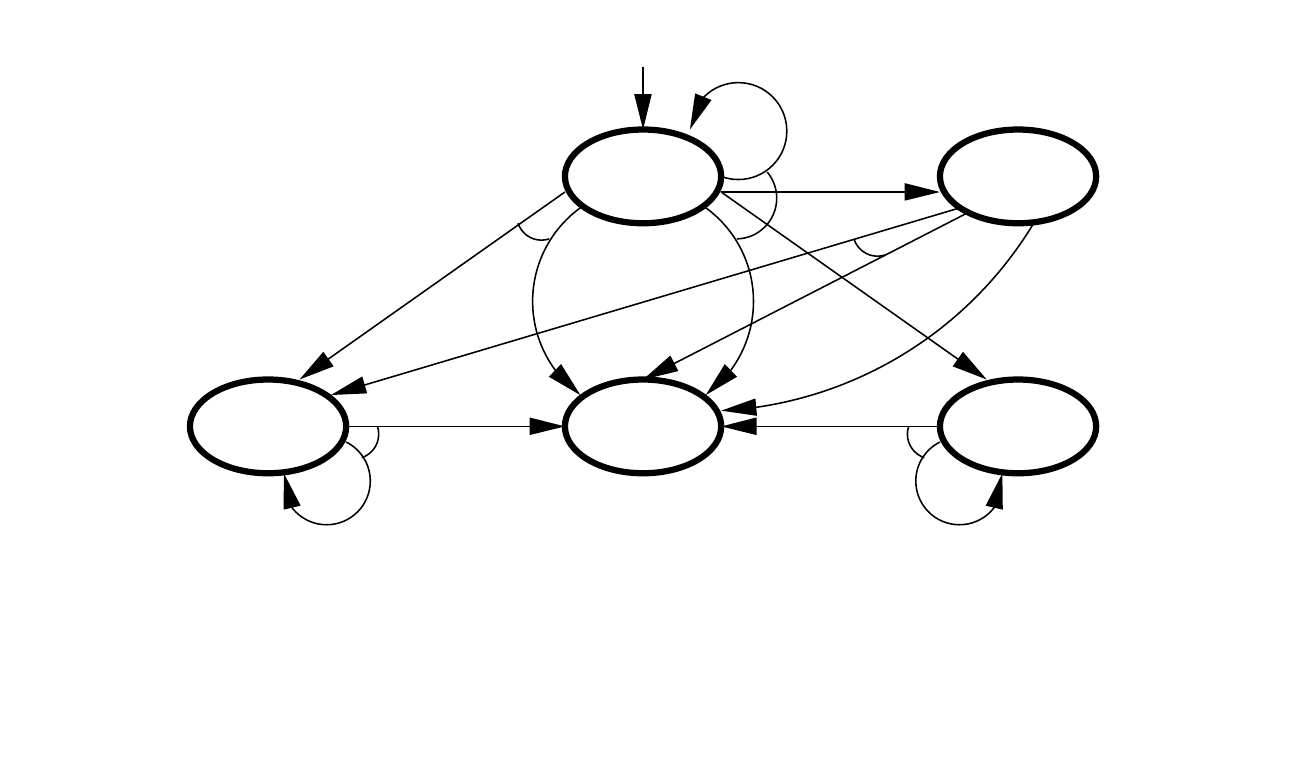}}
\caption{Under-approximations at level 1 and 2 of the difference of
  APAs $N_1$ and $N_2$ from Figure~\ref{fig:c-ex-diff}.}
\label{fig:under-approx}
\end{figure}

Given a state $(s_1,s_2,e,k) \in S$ with $s_2 \ne \bot$ and $e \ne
\epsilon$ and two constraints $\phi_1 \in C(S_1)$ and $\phi_2 \in
C(S_2)$ such that $L_1(s_1,e,\phi_1) \ne \bot$ and $L_2(s_2,e,\phi_2)
\ne \bot$, the constraint $\phi^{B,k}_{12} \in C(S)$ is defined as
follows: $\mu \in Sat(\phi^{B,k}_{12})$ iff
\begin{enumerate}
\item for all $(s'_1,s'_2,c, k') \in S$, if $\mu(s'_1, s'_2,c, k') >0$,
  then $c \in B(s'_1,s'_2) \cup \{\epsilon\}$ and either
  $\textsf{succ}_{s_2,e}(s_1') = \emptyset$, $s'_2 = \bot$ and $k'=1$,
  or $s'_2 = \textsf{succ}_{s_2,e}(s_1')$,
\item the distribution $\mu_1 : s'_1 \mapsto \sum_{c \in
    A\cup\{\epsilon\}, s'_2 \in S_2\cup \{\bot\}, k'\ge 1}
  \mu(s'_1,s'_2,c, k')$ satisfies $\phi_1$, and
\item one of the following holds:
  \begin{enumerate}
  \item there exists $(s'_1, \bot, c, 1)$ such that $\mu(s'_1, \bot, c,
    1) >0$,
  \item the distribution $\mu_2 : s'_2 \mapsto \sum_{c \in
      A\cup\{\epsilon\}, s'_1 \in S_1, k'\ge 1} \mu(s'_1,s'_2,c, k')$
    does not satisfy $\phi_2$, or
  \item $k \ne 1$ and there exists $s'_1 \in S_1$, $s'_2 \in S_2$, $c
    \ne \epsilon$ and $k' < k$ such that $\mu(s'_1,s'_2,c, k')>0$.
  \end{enumerate}
\end{enumerate}
The construction is illustrated in Figure~\ref{fig:under-approx}.

\subsection{Properties}

We already saw in Theorem~\ref{th:over-diff} that $N_1\setminus^* N_2$
is a correct over-approxi\-ma\-tion of the difference of $N_1$ by $N_2$ in
terms of sets of implementations.  The next theorem shows that,
similarly, all $N_1\setminus^K N_2$ are correct under-approximations.
Moreover, increasing the value of $K$ improves the level of
approximation, and eventually all PAs in $\sem{ N_1}\setminus \sem{
  N_2}$ are caught.  (Hence in a set-theoretic sense, $\lim_{ K\to
  \infty} \sem{ N_1\setminus^K N_2}= \sem{ N_1}\setminus \sem{ N_2}$.)

\begin{thm}
  \label{th:under-diff}
  For all deterministic APAs $N_1$ and $N_2$ in SVNF such that $N_1 \not
  \preceq N_2$:
  \begin{enumerate}
  \item for all $K \in \mathbb{N}$, we have $N_1 \setminus^K N_2 \preceq
    N_1 \setminus^{K+1} N_2$,
  \item for all $K \in \mathbb{N}$, $\impl{N_1 \setminus^{K} N_2}
    \subseteq \impl{N_1} \setminus \impl{N_2}$, and
  \item for all PA $P \in \impl{N_1}\setminus \impl{N_2}$, there exists
    $K \in \mathbb{N}$ such that $P \in \impl{N_1 \setminus^{K} N_2}$.
  \end{enumerate}
\end{thm}

Note that item 3 implies that for all PA $P\in \sem{ N_1}\setminus \sem{
  N_2}$, there is a finite specification capturing $\sem{ N_1}\setminus
\sem{ N_2}$ ``up to'' $P$.  The proof of the theorem is similar to the
one of Theorem~\ref{th:over-diff} (if somewhat more complicated) and
available in appendix.

Using our distance defined in Section~\ref{sec:ref+dist}, we can make
the above convergence result more precise.  We first need a lemma
comparing $N_1\setminus^{ K_1} N_2$ with $N_1\setminus^{ K_2} N_2$ for
$K_1\le K_2$.

\begin{lem}
  \label{lem:under-diff-convergence}
  Let $N_1 = (S_1, A, L_1, AP, V_1, \{s_0^1\})$ and $N_2 = (S_2, A, L_2,
  AP, V_2, \{s_0^2\})$ be two deterministic APAs in SVNF such that $N_1
  \not \preceq N_2$. Let $1 \le K_1 \le K_2$ be integers. Then $d(N_1
  \setminus^{K_2} N_2,N_1 \setminus^{K_1} N_2) \le \lambda^{K_1}$.
\end{lem}

\proof
  Let $N_1 \setminus^{K_i} N_2 = N^i = (S^i, A, L^i, AP, V^i, T_0^i)$.
  We first remark that for all $(s_1,s_2,e) \in S_1 \times (S_2\cup
  \bot) \times (A \cup \epsilon)$ and for all $k \le K_1$, the distance
  between the states $(s_1, s_2, e, k)^1 \in S^1$ and $(s_1, s_2, e,
  k)^2 \in S^2$ is $0$. Indeed, if $k$ is the same in both states, then
  they are identical by construction.

  We now prove by induction on $1\le k_1 \le K_1$ and $k_1 \le k_2 \le
  K_2$ that
  \begin{equation*}
    d((s_1,s_2,e,k_2)^2,(s_1,s_2,e,k_1)^1) \le \lambda^{k_1}:
  \end{equation*}
  \begin{description}
  \item[\bf Base case: $k_1 = 1$]  By construction, $t_1 =
    (s_1,s_2,e,k_1)^1$ and $t_2 = (s_1,s_2,e,k_2)^2$ have the same
    outgoing transitions. The only distinction is in the constraints
    $\phi_{12}^{B,1}$ and $\phi_{12}^{B,k_2}$ when $e \in
    B_{\{c,e,f\}}(s_1,s_2)$.  Thus, $t_1$ and $t_2$ are compatible, and
    \begin{equation*}
      d(t_2,t_1) = \max
      \begin{cases}
        \displaystyle %
        \max_{ a,\phi' : L^2(t_2,a,\phi') \ne \bot\,} \min_{ \phi :
          L^1(t_1,a,\phi) \ne \bot} \lambda D_{N^2,N^1}(\phi',\phi,d)
        \\
        \displaystyle %
        \max_{ a,\phi : L^1(t_1,a,\phi) = \top\,} \min_{ \phi' :
          L^2(t_2,a,\phi') = \top} \lambda D_{N^2,N^1}(\phi',\phi,d)
      \end{cases}
    \end{equation*}
    Moreover, we know by construction that $D_{N^2,N^1}(\phi',\phi,d)
    \le 1$ for all $\phi'$ and $\phi$. As a consequence, $d(t_2,t_1) \le
    \lambda = \lambda^{k_1}$.

  \item[\bf Induction] Let $t_1 = (s_1,s_2,e,k_1)^1$ and $t_2 =
    (s_1,s_2,e,k_2)^2$, with $1< k_1 \le k_2$. Again, if $e \notin
    B_c(s_1,s_2) \cup B_e(s_1,s_2) \cup B_f(s_1,s_2)$, then $t_1$ and
    $t_2$ are identical by construction and the result holds. Otherwise,
    the pair of constraints maximizing the distance will be
    constraints $\phi_{12}^{B,k_1} \in C(S^1)$ and $\phi_{12}^{B,k_2}
    \in C(S^2)$. Assume that $d((s_1,s_2,e,k'_2)^2,(s_1,s_2,e,k'_1)^1)
    \le \lambda^{k'_1}$ for all $k'_1 < k_1$ and $k'_1 \le k'_2 \le
    K_2$. By definition, we have
    \begin{align*}
      &\qquad D_{N^2,N^1}(\phi_{12}^{B,k_2}, \phi_{12}^{B,k_1},d) = \\
      &\qquad\qquad\qquad \sup_{\mu_2 \in Sat(\phi_{12}^{B,k_2})\,} \inf_{\delta \in
        \rd(\mu_2,\phi_{12}^{B,k_1})\,} \sum_{t'_2,t'_1 \in S^2\times S^1}
      \mu_2(t'_2)\delta(t'_2,t'_1) d(t'_2,t'_1)
    \end{align*}
    Consider the function $\delta : S^2 \times S^1 \rightarrow [0,1]$
    such that
    \begin{equation*}
      \qquad \delta((s'_1,s'_2,f,k'_2),(s''_1,s''_2,f',k'_1)) =
      \begin{cases}
          1 & \text{if } s'_1 = s''_1 \land s'_2 = s''_2
          \land f' = f \\
          & \qquad \qquad \land\, k'_1 = k'_2 \land k'_2 < k_1  \\
          1 & \text{if } s'_1 = s''_1 \land s'_2 = s''_2 \land f' =
          f \\
          & \qquad \qquad \land\, k'_1 = k_1 -1 \land  k_1 \le k'_2 \\
          0 & \text{otherwise }
      \end{cases}
    \end{equation*}
    Let $\mu_2 \in Sat(\phi_{12}^{B,k_2})$. One can verify that $\delta
    \in \rd(\mu_2, \phi_{12}^{B,k_1})$ as follows:\medskip

    \begin{enumerate}
    \item Let $t'_2 = (s'_1,s'_2,f,k'_2)$ be such that
      $\mu_2(t'_2)>0$. By definition, we always have $\sum_{t'_1 \in
        S^1} \delta(t'_2,t'_1) = 1$.

    \item $\delta$ preserves all the conditions for satisfying
      $\phi_{12}^{B,k_2}$. In particular, all states $t'_2 =
      (s'_1,s'_2,f,k'_2)^2$ such that $k'_2 < k_2$ are redistributed to
      states $(s'_1,s'_2,f,k'_1)^1$ with $k'_1 < k_1$. As a consequence,
      the distribution $\mu_1 : t'_1 \mapsto \sum_{t'_2 \in S^2}
      \mu_2(t'_2) \delta(t'_2,t'_1)$ satisfies $\phi_{12}^{B,k_1}$.
    \end{enumerate}
    
    \noindent As a consequence, for all $\mu_2 \in Sat(\phi_{12}^{B,k_2})$, we
    have
    \begin{align*}
      \hspace*{6em} & \hspace*{-2em} \inf_{\delta \in
        \rd(\mu_2,\phi_{12}^{B,k_1})\,} \sum_{t'_2,t'_1
        \in S^2\times S^1} \mu_2(t'_2)\delta(t'_2,t'_1) d(t'_2,t'_1) \\
      &\le \sum_{\substack{(s'_1,s'_2,f,k'_2) \in S^2 \\ k'_2 < k_1}}
      \mu_2(s'_1,s'_2,f,k'_2)
      d((s'_1,s'_2,f,k'_2)^2,(s'_1,s'_2,f,k'_2)^1) \\
      & \qquad + \sum_{\substack{(s'_1,s'_2,f,k'_2) \in S^2 \\ k_1 \le
          k'_2}} \mu_2(s'_1,s'_2,f,k'_2)
      d((s'_1,s'_2,f,k'_2)^2,(s'_1,s'_2,f,k_1-1)^1) \\
      \intertext{\pagebreak}
      &\le \sum_{\substack{(s'_1,s'_2,f,k'_2) \in S^2 \\ k_1 \le k'_2}}
      \mu_2(s'_1,s'_2,f,k'_2)
      d((s'_1,s'_2,f,k'_2)^2,(s'_1,s'_2,f,k_1-1)^1) \\
      &\le \sum_{\substack{(s'_1,s'_2,f,k'_2) \in S^2 \\ k_1 \le k'_2}}
      \mu_2(s'_1,s'_2,f,k'_2) \lambda^{k_1-1} \le \lambda^{k_1 - 1}
    \end{align*}
    (the next-to-last step by induction).
    Since this is true for all $\mu_2 \in Sat(\phi_{12}^{B,k_2})$, we
    have $D_{N^2,N^1}(\phi_{12}^{B,k_2},\phi_{12}^{B,k_1},d) \le
    \lambda^{k_1-1}$.
    Finally, we have $d(t_2,t_1) \le \lambda \lambda^{k_1 -1} =
    \lambda^{k}$, which proves the induction.
  \end{description}

  \noindent For any state $t_0^2 = (s_0^1,s_0^2,e,K_2) \in T_0^2$, there exists a
  state $t_0^1 = (s_0^1,s_0^2,e,K_1) \in T_0^1$ such that $d(t_0^2,
  t_0^1) \le \lambda^{K_1}$. As a consequence, we have $d(N_1
  \setminus^{K_2} N_2,N_1 \setminus^{K_1} N_2) \le \lambda^{K_1}$.  \qed

The next proposition then shows that the speed of convergence is
exponential in $K$; hence in practice, $K$ will typically not need to be
very large.

\begin{prop}
  \label{pr:underapprox-speed}
  Let $N_1$ and $N_2$ be two deterministic APAs in SVNF such that $N_1
  \not \preceq N_2$, and let $K\in \mathbb{N}$.  Then $d_t( \sem{ N_1}
  \setminus \sem{ N_2}, \sem{ N_1\setminus^K N_2})\le \lambda^K( 1-
  \lambda)^{ -1}$.
\end{prop}

\proof
  By Lemma~\ref{lem:under-diff-convergence}, we know that $d(
  N_1\setminus^{ L+ 1} N_2, N_1\setminus^L N_2)\le \lambda^L$ for each
  $L$, hence also $d_t( \sem{ N_1\setminus^{ L+ 1} N_2}, \sem{
    N_1\setminus^L N_2})\le \lambda^L$ for each $L$ by
  Proposition~\ref{th:distances}.  Applying the triangle inequality and
  continuity of $d_t$, we see that
  \begin{align*}
    d_t( \sem{ N_1}\setminus \sem{ N_2}, \sem{ N_1\setminus^K N_2})
    & \le d_t(\sem{N_1}\setminus\sem{N_2},\sem{N_1\setminus^{K+1}N_2})\\
    & \qquad \qquad + d_t( \sem{ N_1\setminus^{ K+ 1} N_2}, \sem{
      N_1\setminus^{K} N_2}) \\
    &\le \lim_{i\rightarrow \infty}
    d_t(\sem{N_1}\setminus\sem{N_2},\sem{N_1\setminus^{K+i}N_2})\\
    & \qquad \qquad + \sum_{ i= 0}^\infty d_t( \sem{ N_1\setminus^{ K+
        i+ 1} N_2}, \sem{
      N_1\setminus^{ K+ i} N_2}) \\
    &\le \sum_{ i= 0}^\infty \lambda^{ K+ i}= \frac{ \lambda^K}{ 1-
      \lambda}\rlap{\hbox to 170 pt{\hfill\qEd}}
  \end{align*}\medskip

\noindent For the actual application on hand however, the particular accumulating
distance $d$ we have introduced in Section~\ref{sec:ref+dist} may have
limited interest, especially considering that one has to fix a
discounting factor for actually calculating it.  What is more
interesting are results of a \emph{topological} nature which abstract
away from the particular distance used and apply to all distances which
are \emph{topologically equivalent} to $d$.  The results we present
below are of this nature.

It can be shown, cf.~\cite{DBLP:journals/jlp/ThraneFL10}, that
accumulating distances for different choices of $\lambda$ are
topologically equivalent (indeed, even Lipschitz equivalent), hence the
particular choice of discounting factor is not important.  Also some
other system distances are Lipschitz equivalent to the accumulating one,
in particular the so-called \emph{point-wise} and \emph{maximum-lead}
ones, see again~\cite{DBLP:journals/jlp/ThraneFL10}.

\begin{thm}
  \label{th:conv-distances}\label{th:main}
  Let $N_1$ and $N_2$ be two deterministic APAs in SVNF such that $N_1
  \not \preceq N_2$.
  \begin{enumerate}
  \item The sequence $(N_1 \setminus^K N_2)_{K\in \mathbb{N}}$ converges
    in the distance $d$, and $\lim_{K \rightarrow \infty} d(N_1
    \setminus^{*} N_2, N_1 \setminus^K N_2) = 0$.
  \item The sequence $(\impl{N_1 \setminus^K N_2})_{K\in \mathbb{N}}$
    converges in the distance $d_t$, and $\lim_{K \rightarrow \infty}
    d_t( \impl{N_1} \setminus \impl{N_2}, \impl{N_1 \setminus^K N_2}) =
    0$.
  \end{enumerate}
\end{thm}

\proof
  Let $N_1 = (S_1, A, L_1, AP, V_1, \{s_0^1\})$ and $N_2 = (S_2, A, L_2,
  AP, V_2, \{s_0^2\})$ be two deterministic APAs in SVNF such that $N_1
  \not \preceq N_2$.

  \medskip
  \noindent {\bf 1.} The proof of the convergence of both sequences
  $(N_1 \setminus^K N_2)_K$ and $(\impl{N_1 \setminus^K N_2})_K$ is done
  as follows.  Let $\epsilon >0$. Since $\lambda < 1$, there exists $K
  \in \mathbb{N}$ such that $\lambda^{K} < \epsilon$. As a consequence,
  by Lemma~\ref{lem:under-diff-convergence}, we have that for all $K
  \le K_1 \le K_2$,
  \begin{equation*}
    d(N_1 \setminus^{K_2} N_2,N_1 \setminus^{K_1} N_2)
    \le \lambda^{K_1} \le \lambda^{K} < \epsilon.
  \end{equation*}
  The sequence $(N_1 \setminus^{K} N_2)_K$ is thus \emph{bi-Cauchy}
  (i.e.~both forward-Cauchy and backwards-Cauchy) in the sense
  of~\cite{DBLP:journals/tcs/BonsangueBR98}.  Hence, because of
  Proposition~\ref{th:distances}, the sequence (of sets of PA) $( \sem{
    N_1\setminus^K N_2})_K$ is also bi-Cauchy.  The other two items show
  that they converge.

  \medskip
  \noindent {\bf 2.} Theorem~\ref{th:under-diff} shows that the sequence
  $( \sem{ N_1\setminus^K N_2})_K$ converges in a set-theoretic sense
  (as a direct limit), and establishes $\lim_{ K\to \infty} \sem{
    N_1\setminus^K N_2}= \sem{ N_1}\setminus \sem{ N_2}$.  Hence $d_t(
  \sem{ N_1}\setminus \sem{ N_2},\break \lim_{ K\to \infty} \sem{
    N_1\setminus^K N_2})= 0$, and by continuity of $d_t$, $\lim_{K
    \rightarrow \infty} d_t( \impl{N_1} \setminus \impl{N_2}, \impl{N_1
    \setminus^K N_2}) = 0$.

  \medskip
  \noindent {\bf 3.} Finally, we prove that $\lim_{K \rightarrow \infty}
  d(N_1 \setminus^{*} N_2, N_1 \setminus^K N_2) = 0$. This proof is very
  similar to the proof of Lemma~\ref{lem:under-diff-convergence} above:
  we can show that the distance between $N_1 \setminus^{*} N_2$ and $N_1
  \setminus^{K} N_2$ is bounded as follows:
  \begin{equation*}
    d(N_1 \setminus^{*} N_2,N_1 \setminus^{K} N_2) \le \lambda^{K}
  \end{equation*}
  Let $N_1 \setminus^{K} N_2 = N^K = (S^K, A, L^K, AP, V^K, T_0^K)$,
  $N_1 \setminus^{*} N_2 = N^{*} = (S^{*}, A, L^{*}, AP, V^{*},
  T_0^{*})$.  We start by proving by induction on $1 \le k \le K$ that
  for all $(s_1,s_2,e) \in S_1 \times (S_2\cup \bot) \times (A \cup
  \epsilon)$, we have $d((s_1,s_2,e)^{*},(s_1,s_2,e,k)) \le
  \lambda^{k}$. The only difference with the proof of
  Lemma~\ref{lem:under-diff-convergence} is in the choice of the
  function $\delta : S^{*} \times S^{K} \rightarrow [0,1]$ in the
  induction part. Here, we choose $\delta$ as follows:
  \begin{equation*}
    \delta((s'_1,s'_2,f),(s''_1,s''_2,f',k')) =
    \begin{cases}
        1 & \text{if } s'_1 = s''_1 \land s'_2 = s''_2 \land f' =
        f \land k' = k-1 \\
        0 & \text{otherwise }
    \end{cases}
  \end{equation*}
  The rest of the proof is identical, and we obtain that for all $1 \le
  k \le K$ and for all $(s_1,s_2,e) \in S_1 \times (S_2\cup \bot) \times
  (A \cup \epsilon)$, we have $d((s_1,s_2,e)^{*},(s_1,s_2,e,k)) \le
  \lambda^{k}$. In particular, this is also true for initial states.  As
  a consequence, for all states $t_0^{*} = (s_0^{2},s_0^1,e) \in
  T_0^{*}$, there exists a state $t_0^K = (s_0^1,s_0^2,e,K) \in T_0^K$
  such that $d(t_0^{*}, t_0^K) \le \lambda^{K}$, hence we have $d(N_1
  \setminus^{*} N_2,N_1 \setminus^{K} N_2) \le \lambda^{K}$, so that
  $\lim_{K \rightarrow \infty} d(N_1 \setminus^{*} N_2, N_1 \setminus^K
  N_2) = 0$.
\qed

Recall that as $d$ and $d_t$ are not metrics, but only (asymmetric)
pseudometrics (i.e.~hemi-metrics), the above sequences may have more
than one limit; hence the particular formulation.  The theorem's
statements are topological, as they only allude to convergence of
sequences and distance $0$; topologically equivalent distances obey
precisely the property of having the same convergence behavior and the
same kernel, cf.~\cite{books/AliprantisB07}.

The next corollary, which is easily proven from the above theorem by
noticing that its first part implies that also $\lim_{K \rightarrow
  \infty} d_t(\sem{N_1 \setminus^{*} N_2}, \sem{N_1 \setminus^K N_2}) =
0$, shows what we mentioned already at the end of
Section~\ref{sec:over-approx}: with respect to the distance $d$,
$N_1\setminus^* N_2$ is a best possible over-approximation of $\sem{
  N_1}\setminus \sem{ N_2}$.

\begin{cor}
  \label{co:over-approx-best}
  Let $N_1$ and $N_2$ be two deterministic APAs in SVNF such that $N_1
  \not \preceq N_2$.  Then $d_t(\impl{N_1 \setminus^{*} N_2}, \impl{N_1}
  \setminus \impl{N_2}) = 0$.
\end{cor}

Again, as $d_t$ is not a metric, the distance being zero does not imply
that the sets $\sem{ N_1\setminus^* N_2}$ and $\sem{ N_1}\setminus \sem{
  N_2}$ are equal; it merely means that they are
\emph{indistinguishable} by the distance $d_t$, or infinitesimally close
to each other.

\section{Counter-Example Generation}
\label{sec:algo}

Here we show how some techniques similar to the ones we have introduced
can be used to generate \emph{one} counterexample to a failed refinement
$N_1\not\preceq N_2$.  Note that when we compute the approximating
differences $N_1\setminus^* N_2$ and $N_1\setminus^K N_2$, we are in
principle generating (approximations to) the set of \emph{all}
counterexamples, hence what we do in Section~\ref{sec:diff} is much more
general than what we will present below.  Generating only \emph{one}
counterexample may still be interesting however, as it is somewhat
easier than computing the differences $N_1\setminus^* N_2$,
$N_1\setminus^K N_2$ and is all that is needed in a CEGAR approach.

First remark that Definition~\ref{def:refinement} can be trivially
turned into an algorithm for checking refinement. Let
$N_1=(S_1,A,L_1,AP,V_1,\{s_0^1\})$ and
$N_2=(S_2,A,L_2,AP,V_2,\{s_0^2\})$ be two deterministic APAs in
SVNF. Consider the initial relation $\rel_0 = S_1 \times S_2$. Compute
$\rel_{k+1}$ by removing all pairs of states not satisfying
Definition~\ref{def:refinement} for $\rel_k$.  The sequence $(\rel_n)_{n
  \in \mathbb{N}}$ is then strictly decreasing and converges to a
fixed point within a finite number of steps $K \le |S_1 \times S_2|$. This
fixed point $\rel_K$ coincides with the maximal refinement relation $\rel$
between $N_1$ and $N_2$. Let the index of this fixed point be denoted with
$\ind(\rel) = K$; hence $\ind_{\rel}(s_1,s_2) = \min(\max(\{k \st
(s_1,s_2)\in \rel_k\}),K)$.

We now observe that if a pair of states $(s_1,s_2)$ is removed from the
relation $\rel$ by case $3$, then we need to keep track of the actions
that lead to this removal in order to use them in our
counterexample. Whenever a pair of states is in cases 3.a, 3.b, 3.d or
3.e, we have that $\ind_{\rel}(s_1,s_2)= 0$ and the counterexample can
be easily produced by allowing or disallowing the corresponding
transitions from $N_1$ and $N_2$. Cases 3.c and 3.f play a different
role: due to the fact that they exploit distributions, they are the only
cases in which refinement can be broken by using its \emph{recursive}
axiom. In these cases, producing a counterexample can be done in two
ways: either by using a distribution that does not satisfy the
constraints in $N_2$ (if such a distribution exists, then
$\ind_{\rel}(s_1,s_2) = 0$), or by using a distribution that reaches a
pair of states $(s'_1,s'_2) \notin \rel$. When $0< \ind_{\rel}(s_1,s_2)
< \ind(\rel)$, only the latter is possible. This recursive construction
has disadvantages: it allows us to produce loops that may lead to
incorrect counterexamples. In order to prevent these loops, we propose
to use only those distributions that decrease the value of $\ind$ in
this particular case. The set $\breaking(s_1,s_2)$ defined hereafter
allows us to distinguish the actions for which the value of $\ind$
decreases, hence ensuring (by Lemma~\ref{lem:ind+} below) the
correctness of our counterexample construction.  Let $(s_1,s_2) \in S_1
\times S_2$ be such that $V_1(s_1)= V_2(s_2)$ and $\ind_{\rel}(s_1,s_2)
= k < \ind(\rel)$. We define
\begin{align*}
  \breaking(s_1,s_2) &= \{ a \in A\mid a \in B_{a,b,d,e}(s_1,s_2),
  \text{ or } \\
  &\hspace*{3em} \exists \phi_1 \in C(S_1), \phi_2 \in C(S_2), \mu_1 \in
  Sat(\phi_1): \\
  &\hspace*{3em} L_1(s_1,a,\phi_1) \neq \bot, L_2(s_2,a,\phi_2) \ne \bot,
  \forall \mu_2 \in Sat(\phi_2): \mu_1 \not \leqbox_{\rel_{k}} \mu_2 \}
\end{align*}

Observe that the conditions for $\breaking$ above are exactly the
conditions for removing a pair of states $(s_1,s_2)$ at step $k$ of the
algorithm for computing $\rel$ defined above. Under the assumption that
$V_1(s_1) \subseteq V_2(s_2)$ and $\ind_{\rel}(s_1,s_2) = k <
\ind(\rel)$, we can be sure that the set $\breaking(s_1,s_2)$ is not
empty. Moreover, we have the following lemma.

\begin{lem}
  \label{lem:ind+}
  For all pairs of states $(s_1,s_2)$ in case 3 and for all actions $e
  \in (B_c(s_1,s_2) \cup B_f(s_1,s_2))\cap \breaking(s_1,s_2)$, there
  exist constraints $\phi_1$ and $\phi_2$ such that $L_1(s_1,e,\phi_1)
  \ne \bot$ and $L_2(s_2,e,\phi_2) \ne \bot$ and a distribution $\mu_1
  \in Sat(\phi_1)$ such that
  \begin{enumerate}
  \item $\exists s'_1 \in S_1$ such that $\mu_1(s'_1) >0$ and
    $\textsf{succ}_{s_2,e}(s'_1) = \emptyset$,

  \item $\mu_1^2 : \big( s'_2 \mapsto \sum_{\{s'_1 \in S_1 \st s'_2 =
        \textsf{succ}_{s_2,e}(s'_1)\}} \mu_1(s'_1) \big) \notin
    Sat(\phi_2)$, or

  \item $\exists s'_1\in S_1, s'_2\in S_2$ such that $\mu_1(s'_1)>0,
    s'_2 = \textsf{succ}_{s_2,e}(s'_1)$ and $\ind_{\rel}(s'_1,s'_2) <
    \ind_{\rel}(s_1,s_2)$.
  \end{enumerate}
\end{lem}

\proof
  Let $\rel$ be the maximal refinement relation between $N_1$ and $N_2$
  and let $(s_1,s_2) \in S_1 \times S_2$ such that $(s_1,s_2)$ is in
  case 3, i.e.~$(s_1,s_2) \notin \rel$ and $V_1(s_1) = V_2(s_2)$. Let $e
  \in A$ such that $e \in (B_c(s_1,s_2) \cup B_f(s_1,s_2))\cap
  \breaking(s_1,s_2)$.

  Since $e \in B_c(s_1,s_2) \cup B_f(s_1,s_2)$, there exists $\phi_1 \in
  C(S_1)$ and $\phi_2 \in C(S_2)$ such that either $L_2(s_2,e,\phi_2) =
  \top$ and $L_1(s_1,e,\phi_1) = \top$ or $L_2(s_2,e,\phi_2) = \may$ and
  $L_1(s_1, e, \phi_1) \ne \bot$. As a consequence, since $e \in
  \breaking(s_1,s_2)$, we have that
  \begin{equation}
    \label{eq:eq-1}
    \exists \mu_1 \in Sat(\phi_1), \forall \mu_2 \in Sat(\phi_2),
    \mu_1 \not \leqbox_{\rel_k} \mu_2.
  \end{equation}

  Let $K$ be the smallest index such that $\rel_K = \rel$. By
  construction, we know that $\ind_{\rel}(s_1,s_2) = k <K$,
  i.e.~$(s_1,s_2) \in \rel_k$ and $(s_1,s_2) \notin
  \rel_{k+1}$. Consider the distribution $\mu_1$ given by
  (\ref{eq:eq-1}) above. We have that $\forall \mu_2 \in Sat(\phi_2):
  \forall \text{ corresp.~}\delta: \mu_1 \not \leqbox_{\rel_k}^{\delta}
  \mu_2$. Consider the function $\delta$ such that $\delta(s'_1,s'_2) =
  1$ if $s'_2 = \textsf{succ}_{s_2,e}(s'_1)$ and $0$ otherwise. There
  are several cases.
  \begin{itemize}
  \item If there exists $s'_1 \in S_1$ such that $\mu_1(s'_1) >0$ and
    $\textsf{succ}_{s_2,e}(s'_1) = \emptyset$, then the lemma is proven.

  \item Else, $\delta$ is a correspondence function. Since $\forall
    \mu_2 \in Sat(\phi_2), \mu_1 \not \leqbox_{\rel_k} \mu_2$, we know
    that either (1) $\mu_2 : s'_2 \mapsto \sum_{s'_1 \in S_1}
    \mu_1(s'_1) \delta(s'_1,s'_2)$ does not satisfy $\phi_2$, or (2)
    there exists $s'_1$ and $s'_2$ such that $\mu_1(s'_1) >0$,
    $\delta(s'_1,s'_2)>0$ and $(s'_1,s'_2) \notin \rel_k$.
    \begin{enumerate}
    \item Assume that $\mu_2 : s'_2 \mapsto \sum_{s'_1 \in S_1}
      \mu_1(s'_1) \delta(s'_1,s'_2)$ does not satisfy $\phi_2$. Remark
      that the function $\mu_1^2$ from Lemma~\ref{lem:ind+} is equal to
      $\mu_2$ defined above. As a consequence, $\mu_1^2 \notin \phi_2$.

    \item Otherwise, assume that there exists $s'_1$ and $s'_2$ such
      that $\mu_1(s'_1) >0$, $\delta(s'_1,s'_2)>0$ and $(s'_1,s'_2)
      \notin \rel_k$. Since $(s'_1,s'_2) \notin \rel_k$, we have that
      $\ind_{\rel}(s'_1,s'_2) < k$. As a consequence, there exists
      $s'_1\in S_1, s'_2\in S_2$ such that $\mu_1(s'_1)>0, s'_2 =
      \textsf{succ}_{s_2,e}(s'_1)$ and $\ind_{\rel}(s'_1,s'_2) <
      \ind_{\rel}(s_1,s_2)$. \qed
    \end{enumerate}
  \end{itemize}

\noindent In other words, the above lemma ensures that a pair $(s'_1,
s'_2)$ such that $\ind_{\rel}(s'_1,s'_2) = 0$ can be reached within a
bounded number of transitions for all pairs of states $(s_1,s_2)$ in
case 3. As explained above, this is a prerequisite for the correctness
of the counterexample construction defined hereafter.

We now propose a construction to build counterexamples.  Consider
deterministic APAs $N_1=(S_1,A,L_1,AP,V_1,\{s_0^1\})$ and
$N_2=(S_2,A,L_2,AP,V_2,\{s_0^2\})$ in SVNF such that $N_1 \not \preceq
N_2$. Let $\rel$ be the maximal refinement relation between $N_1$ and
$N_2$.

\begin{table}
  \caption{%
    \label{tab:c-ex}
    Definition of the transition function $L$ in $P$.}
  \begin{tabular}{|l|c|c|l|}
    \hline
    $e \in$ & $N_1, N_2$ & $P$ & Formal Definition of $L$\\ \hline
    
    $B_a(s_1,s_2)$  & 
    \begin{minipage}[c][1.5cm]{0.13\linewidth}
      \raisebox{-1.4cm}{\hspace{-0.2cm}{\scalebox{0.4}{\input{case3a.pdf_t}}}}
    \end{minipage}
    &
    \multirow{2}{*}{\begin{minipage}[c][1.8cm]{0.12\linewidth}
        \raisebox{-1cm}{{\scalebox{0.5}{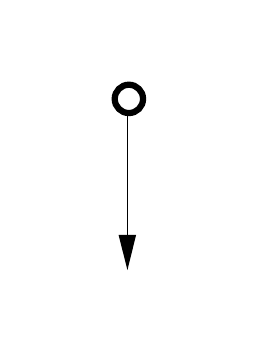}}}
      \end{minipage}}
    & 
    \multirow{2}{*}{\begin{minipage}[c][2cm]{0.6\linewidth} Let
        $\phi_1 \in C(S_1)$ such that $L_1(s_1, e ,\phi_1) \ne \bot$
        and let $\mu_1$ be an arbitrary distribution in
        $Sat(\phi_1)$. Define $L((s_1,s_2),e,\mu_1^{\bot}) = \top$.
      \end{minipage}}
    
    \\\cline{1-2}
    
    $B_b(s_1,s_2)$  & 
    \begin{minipage}[c][1.5cm]{0.13\linewidth}
      \raisebox{-1.4cm}{\hspace{-0.2cm}{\scalebox{0.4}{\input{case3b.pdf_t}}}}
    \end{minipage}
    &
    &

    \\\hline
    
    $B_d(s_1,s_2)$  & 
    \begin{minipage}[c][1.5cm]{0.13\linewidth}
      \raisebox{-1.4cm}{\hspace{-0.2cm}{\scalebox{0.4}{\input{case3d.pdf_t}}}}
    \end{minipage}
    &
    \multirow{2}{*}{\begin{minipage}[c][2cm]{0.12\linewidth}
        \raisebox{-1cm}{{\scalebox{0.5}{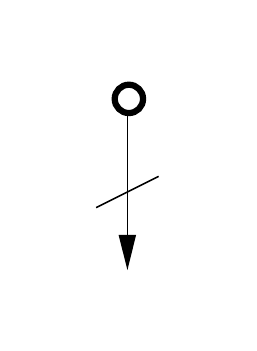}}}
      \end{minipage}}
    &
    \multirow{2}{*}{\begin{minipage}[c][2cm]{0.6\linewidth}
        For all $\mu \in Dist(S)$, let $L((s_1,s_2),e,\mu) = \bot$.
      \end{minipage}}
    
    \\\cline{1-2}
    
    $B_e(s_1,s_2)$  & 
    \begin{minipage}[c][1.5cm]{0.13\linewidth}
      \raisebox{-1.4cm}{\hspace{-0.2cm}{\scalebox{0.4}{\input{case3e.pdf_t}}}}
    \end{minipage}
    &
    &
    \\\hline
    
    $B_c(s_1,s_2)$  & 
    \begin{minipage}[c][1.8cm]{0.13\linewidth}
      \raisebox{-1.4cm}{\hspace{-0.2cm}{\scalebox{0.4}{\input{case3c.pdf_t}}}}
    \end{minipage}
    &
    \multirow{2}{*}{\begin{minipage}[c][2cm]{0.12\linewidth}
        \raisebox{-1.6cm}{{\scalebox{0.5}{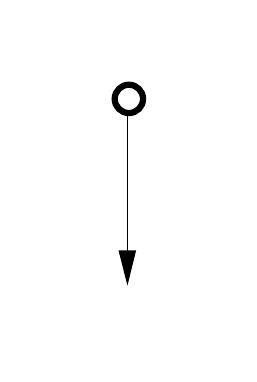}}}
      \end{minipage}}
    & 
    \multirow{2}{*}{\begin{minipage}[c][2.6cm]{0.6\linewidth} Let
        $\phi_1 \in C(S_1)$ and $\phi_2 \in C(S_2)$ such that
        $L_1(s_1, e ,\phi_1) \ne \bot$ and $L_2(s_2,e,\phi_2) \ne
        \bot$. 
        \begin{itemize}
        \item If $e \in \breaking(s_1,s_2)$, then let $\mu_1$ be the
          distribution given in Lemma~\ref{lem:ind+}.
        \item Else, let $\mu_1$ be an arbitrary distribution in
          $Sat(\phi_1)$ such that $\forall \mu_2 \in Sat(\phi_2), \mu_1
          \not \leqbox_{\rel} \mu_2$.
        \end{itemize}
        In both cases, let $L((s_1,s_2),e,\widehat{\mu_1}) = \top$.
      \end{minipage}}
    
    \\\cline{1-2}
    $B_f(s_1,s_2)$
    & 
    \begin{minipage}[c][1.8cm]{0.13\linewidth}
      \raisebox{-1.4cm}{\hspace{-0.2cm}{\scalebox{0.4}{\input{case3f.pdf_t}}}}
    \end{minipage}
    &
    &
    \\\hline
  \end{tabular}
\end{table}

\begin{defi}
  The counterexample $P = (S,A,L,AP,V,s_0)$ is computed as follows:
  \begin{itemize}
  \item $S = S_1 \times (S_2 \cup \{\bot\})$, $s_0 = (s_0^1,s_0^2)$,

  \item $V(s_1,s_2) = v \in 2^{AP}$ such that $V_1(s_1) = \{v\}$ for all
    $(s_1,s_2) \in S$, and

  \item $L$ is defined as follows. Let $(s_1,s_2) \in S$.
    \begin{itemize}
    \item If $(s_1,s_2)$ in case $1$ or $2$ or $s_2 = \bot$, then for
      all $a \in A$ and $\phi_1 \in C(S_1)$ such that $L_1(s_1,a,\phi_1)
      = \top$, let $\mu_1$ be an arbitrary distribution in $Sat(\phi_1)$
      and define $L((s_1,s_2),a,\mu_1^\bot) = \top$ with $\mu_1^\bot \in
      Dist(S)$ such that $\mu_1^\bot(s'_1,s'_2) = \mu_1(s'_1)$ if $s'_2
      = \bot$ and $0$ otherwise.

    \item Else, $(s_1,s_2)$ is in case 3 and $B(s_1,s_2) \ne
      \emptyset$. For all $a \in A \setminus B(s_1,s_2)$ and $\phi_1 \in
      C(S_1)$ such that $L_1(s_1,a,\phi_1) = \top$, let $\mu_1$ be an
      arbitrary distribution in $Sat(\phi_1)$ and let
      $L((s_1,s_2),a,\mu_1^\bot) = \top$, with $\mu_1^\bot$ defined as
      above.

      In addition, for all $e \in B(s_1,s_2)$, let $L((s_1,s_2),e,.)$ be
      defined as in Table~\ref{tab:c-ex}. In the table, given
      constraints $\phi_1 \in C(S_1)$ and $\phi_2 \in C(S_2)$ such that
      $L_1(s_1,e,\phi_1) \ne \bot$ and $L_2(s_2,e,\phi_2) \ne \bot$, and
      a distribution $\mu_1 \in Sat(\phi_1)$, the distribution
      $\widehat{\mu_1} \in Dist(S)$ is defined as follows:
      $\widehat{\mu_1}(s'_1,s'_2) = \mu_1(s_1)$ if $s'_2 =
      \textsf{succ}_{s_2,e}(s'_1)$ or $\textsf{succ}_{s_2,e}(s'_1) =
      \emptyset$ and $s'_2 = \bot$, and $0$ otherwise.
    \end{itemize}
  \end{itemize}
\end{defi}

\begin{thm}
  \label{th:c-ex}
  The counterexample PA $P$ defined above is such that $P \sat N_1$ and
  $P \not \sat N_2$.
\end{thm}

The proof of this theorem is similar to the one of
Theorem~\ref{th:over-diff} and available in appendix.

\section{Conclusion}

We have in this paper added an important aspect to the specification
theory of Abstract Probabilistic Automata, in that we have shown how to
exhaustively characterize the \emph{difference} between two
deterministic specifications.  In a stepwise refinement methodology,
difference is an important tool to gauge refinement failures.

We have also introduced a notion of \emph{discounted distance} between
specifications which can be used as another measure for how far one
specification is from being a refinement of another.  Using this
distance, we were able to show that our sequence of under-approximations
converges, semantically, to the real difference of sets of
implementations, and that our over-approximation is infinitesimally
close to the real difference.

There are many different ways to measure distances between
implementations and specifications, allowing to put the focus on either
transient or steady-state behavior. In this paper we have chosen one
specific discounted distance, placing the focus on transient behavior.
Apart from the fact that this can indeed be a useful distance in
practice, we remark that the convergence results about our under- and
over-approximations are topological in nature and hence apply with
respect to all distances which are topologically equivalent to the
specific one used here, typically discounted distances. Although the
results presented in the paper do not hold in general for the
accumulating (undiscounted) distance, there are other notions of
distances that are more relevant for steady-state behavior,
e.g.~limit-average. Whether our results hold in this setting remains
future work.

We also remark that we have shown that it is not more difficult to
compute the difference of two APAs than to check for their refinement.
Hence if a refinement failure is detected (for example by using the
methods in the APAC tool~\cite{QEST11}), it is not difficult to also
compute the difference for assessing the reason for refinement failure.
For the class of APAs with polynomial constraints, which is the one
implemented in APAC, refinement checking can be done in time quadratic
in the number of states and doubly-exponential in the number of
constraints~\cite{journals/iandc/DelahayeKLLPSW13}; in APAC, the
Z3~solver~\cite{DBLP:conf/tacas/MouraB08} is used for operations on
constraints.

One limitation of our approach is the use of \emph{deterministic} APAs.
Even though deterministic specifications are generally considered to
suffice from a modeling point of view~\cite{DBLP:conf/avmfss/Larsen89},
non-determinism may be introduced for example when composing
specifications.  Indeed, our constructions themselves introduce
non-determinism: for deterministic APAs $N_1$, $N_2$, both
$N_1\setminus^* N_2$ and $N_1\setminus^K N_2$ may be non-deterministic.
Hence it is of interest to extend our approach to non-deterministic
specifications.  The problem here is, however, that for
non-deterministic specifications, the relation between refinement and
inclusion of sets of implementations $N_1\preceq N_2\Longleftrightarrow
\impl{ N_1}\subseteq \impl{ N_2}$ breaks: we may well have
$N_1\not\preceq N_2$ but $\impl{ N_1}\subseteq \impl{ N_2}$,
cf.~\cite{VMCAI11}.  So the technique we have used in this paper to
compute differences will not work for non-deterministic APAs, and
techniques based on \emph{thorough refinement} will have to be
used. 

As a last note, we wish to compare our approach of difference between
APA specifications with the use of \emph{counterexamples} in
probabilistic model checking.  Counterexample generation is studied in a
number of
papers~\cite{%
DBLP:journals/tse/AljazzarL10,%
DBLP:journals/tse/HanKD09,%
DBLP:conf/vmcai/WimmerBB09,%
DBLP:conf/hvc/AndresDR08,%
DBLP:conf/atva/JansenAKWKB11,%
DBLP:conf/concur/SchmalzVV09,%
DBLP:conf/cav/HermannsWZ08,%
DBLP:conf/tacas/WimmerJABK12,%
DBLP:journals/tocl/ChadhaV10,%
DBLP:conf/cav/KomuravelliPC12},
typically with the purpose of embedding it into a procedure of
counterexample guided abstraction refinement (CEGAR).  The focus
typically is on generation of \emph{one} particular counterexample to
refinement, which can then be used to adapt the abstraction accordingly.

In contrast, although we propose a construction for building single
counter-examples, our main focus is on computing APA difference,
i.e.~generating a representation of \emph{all counterexamples}. Our goal
is not to refine abstractions at \emph{system} level, using
counterexamples, but to assess \emph{specifications}.  This is, then,
the reason why we want to compute all counterexamples instead of only
one.  Our work is hence supplementary and orthogonal to the CEGAR-type
use of counterexamples: CEGAR procedures can be used also to refine APA
specifications, but only our difference can assess the precise
distinction between specifications.

\bibliographystyle{plain}
\bibliography{mybib}

\newpage
\appendix

\section*{Appendix: Proof of Theorem~\ref{th:under-diff}}

\proof[Proof of Theorem~\ref{th:under-diff}]
  For the first claim, consider the relation $\rel \subseteq (S_1 \times
  (S_2 \cup \{\bot\})\times (A \cup \{\epsilon\})\times \{1, \ldots,
  K\}) \times (S_1 \times (S_2 \cup \{\bot\})\times (A \cup
  \{\epsilon\})\times \{1, \ldots, K+1\})$ such that $\rel = \{((s_0^1,
  s_0^2, e, K),(s_0^1, s_0^2, e, K+1)) \st e \in B(s_0^1,s_0^2)\} \cup
  \rel_{\mathsf{id}}$, where $\rel_{\mathsf{id}}$ denotes the identity
  relation. One can verify that, by construction, $\rel$ is a refinement
  relation witnessing $N_1 \setminus^K N_2 \preceq N_1 \setminus^{K+1}
  N_2$.

  Let $N_1=(S_1,A,L_1,AP,V_1,\{s_0^1\})$ and
  $N_2=(S_2,A,L_2,AP,V_2,\{s_0^2\})$ be deterministic APAs in single
  valuation normal form such that $N_1\not\preceq N_2$. Let $\rel$ be
  the maximal refinement relation between $N_1$ and $N_2$.

  \medskip
  \noindent {\bf 1.} We first prove that for all $K \in \mathbb{N}$,
  $\impl{N_1 \setminus^{K} N_2} \subseteq \impl{N_1} \setminus
  \impl{N_2}$.  If $V_1(s_0^1) \ne V_2(s_0^2)$, then for all $K\in
  \mathbb{N}$, we have $N_1 \setminus^{K} N_2 = N_1$ and the result
  holds.

  Otherwise, assume that $(s_0^1, s_0^2)$ is in case 3 and let $K \in
  \mathbb{N}$. We have $N_1 \setminus^K N_2 = (S,A,L,AP,V,S_0^K)$
  defined as in Section~\ref{sec:under-approx}. Let $P = (S_P, A, L_P,
  AP, V_P, s_0^P)$ be a PA such that $P \sat N_1 \setminus^{K} N_2$. Let
  $\rel^{\setminus} \subseteq S_P \times S$ be the associated
  satisfaction relation and let $f \in B(s_0^1, s_0^2)$ be such that
  $s_0^P \rel^{\setminus} (s_0^1,s_0^2,f,K)$. We show that $P \sat N_1$
  and $P \not \sat N_2$.

  We start by proving that $P \sat N_1$.  Consider the relation $\rel_1
  \subseteq S_P \times S_1$ such that $p \rel_1 s_1 \iff \exists s_2 \in
  (S_2\cup\{\bot\}), \exists e \in (A\cup\{\epsilon\}), \exists n \le K$
  s.t. $p \rel^{\setminus} (s_1,s_2,e, n)$. We prove that $\rel_1$ is a
  satisfaction relation.  Let $p, s_1, s_2, e, n$ such that $p \rel_1
  s_1$ and $p \rel^{\setminus} (s_1,s_2,e,n)$.
  \begin{itemize}
  \item By construction, we have $V_P(p) \in V((s_1,s_2,e,n)) =
    V_1(s_1)$.

  \item Let $a \in A$ and $\mu_P \in Dist(S_P)$ be such that
    $L_P(p,a,\mu_P) = \top$. By $\rel^{\setminus}$, there exists $\phi
    \in C(S)$ such that $L((s_1,s_2,e,n),a,\phi) \ne \bot$ and there
    exists $\mu \in Sat(\phi)$ such that $\mu_P
    \leqbox_{\rel^{\setminus}} \mu$.

    If $s_2 = \bot$ or $e = \epsilon$ or $a \ne e$, then by construction
    of $N_1 \setminus^K N_2$, there exists $\phi_1 \in C(S_1)$ such that
    $\phi = \phi_1^{\bot}$ and $L_1(s_1,a,\phi_1) \ne \bot$. As a
    consequence, the distribution $\mu\downarrow_1 : s'_1 \mapsto
    \mu(s'_1,\bot,\epsilon, 1)$ is in $Sat(\phi_1)$ and it follows that
    $\mu_P \leqbox_{\rel_1} \mu\downarrow_1$.

    Otherwise, assume that $s_2 \in S_2$, $e \in A$ and $a = e$. There
    are several cases.
    \begin{itemize}
    \item If $e \in B_a(s_1,s_2) \cup B_b(s_1,s_2)$, then by
      construction of $N_1 \setminus^K N_2$, there exists $ \phi_1 \in
      C(S_1)$ such that $L_1(s_1,e,\phi_1) \ne \bot$ and $\phi =
      \phi_1^{\bot}$. As above, we thus have $\mu_P \leqbox_{\rel_1}
      \mu\downarrow_1$.

    \item Else, if $e \in B_e(s_1,s_2)$, then there exists $\phi_1 \in
      C(S_1)$ and $\phi_2 \in C(S_2)$ such that $L_1(s_1,e,\phi_1) ={}?$
      and $L_2(s_2,e,\phi_2) = \top$. Moreover, $\phi$ is of the form
      $\phi_{12}^B$, and $\mu' \in Sat(\phi_{12}^B)$ implies that the
      distribution 
\[\mu_1 : s'_1 \mapsto \sum_{c \in A\cup\{\epsilon\}, s'_2 \in
        S_2\cup \{\bot\}, k'\ge 1} \mu(s'_1,s'_2,c, k')\]
      satisfies
      $\phi_1$. Let $\delta_1 : S_P \rightarrow (S_1 \rightarrow [0,1])$
      be such that $\delta_1(p')(s'_1) = 1$ if $\mu_P(p') >0$ and $s'_1
      = \textsf{succ}_{s_1,e}(p')$ and $0$ otherwise. By construction,
      $\delta_1$ is a correspondence function and we have $\mu_P
      \delta_1 = \mu_1$.

      Thus there exists $\mu_1 \in Sat(\phi_1)$ such that $\mu_P
      \leqbox_{\rel_1} \mu_1$.

    \item Finally, if $e \in B_c(s_1,s_2) \cup B_f(s_1,s_2)$, then there
      exists $\phi_1 \in C(S_1)$ such that $L(s_1,e,\phi_1) \ne \bot$,
      and either $\phi = \phi_1^{\bot}$ or $\phi = \phi_{12}^B$ as in
      the case above. In both cases, as proven before, there exists
      $\mu_1 \in Sat(\phi_1)$ such that $\mu_P \leqbox_{\rel_1} \mu_1$.
    \end{itemize}

  \item Let $a \in A$ and $\phi_1 \in C(S_1)$ such that
    $L_1(s_1,a,\phi_1) = \top$.
  
    If $s_2 = \bot$ or $e = \epsilon$ or $a \ne e$, then by construction
    of $N_1 \setminus^K N_2$, the constraint $\phi_1^{\bot}$ is such
    that $L((s_1,s_2,e,n),a,\phi_1^{\bot}) = \top$. As a consequence,
    there exists a distribution $\mu_P \in Dist(S_P)$ such that
    $L_P(p,a,\mu_P) = \top$ and there exists $\mu \in
    Sat(\phi_1^{\bot})$ such that $\mu_P \leqbox_{\rel^{\setminus}}
    \mu$.  Moreover, by construction of $\phi_1^{\bot}$, the
    distribution $\mu\downarrow_1 : s'_1 \mapsto \mu(s'_1,\bot,\epsilon,
    1)$ is in $Sat(\phi_1)$ and it follows that $\mu_P \leqbox_{\rel_1}
    \mu\downarrow_1$.

    Otherwise, assume that $s_2 \in S_2$, $e \in A$ and $a = e$. Since
    $L_1(s_1,a,\phi_1) = \top$, $(s_1,s_2)$ can only be in cases $3.a,
    3.c$ or $3.f$. As a consequence, $e \in B_a(s_1,s_2) \cup
    B_c(s_1,s_2) \cup B_f(s_1,s_2)$. By construction, in all of these
    cases, we have $L((s_1,s_2,e,n),a,\phi_1^{\bot}) = \top$. Thus,
    there exists a distribution $\mu_P \in Dist(S_P)$ such that
    $L_P(p,a,\mu_P) = \top$ and there exists $\mu \in
    Sat(\phi_1^{\bot})$ such that $\mu_P \leqbox_{\rel^{\setminus}}
    \mu$. As above, it follows that $\mu_P \leqbox_{\rel_1}
    \mu\downarrow_1$.
  \end{itemize}

  Finally, $\rel_1$ is a satisfaction relation. Moreover, by hypothesis,
  we have $s_0^P \rel^{\setminus} (s_0^1,s_0^2, f, K)$, thus $s^P_0
  \rel_1 s_0^1$ and $P \sat N_1$.

  We now prove that $P \not \sat N_2$. Assume the contrary and let
  $\rel_2 \subseteq S_P \times S_2$ be the smallest satisfaction
  relation witnessing $P \sat N_2$ (i.e. containing only reachable
  states). We prove the following by induction on the value of $n$, for
  $1 \le n \le K$: $\forall p \in S_P, s_2 \in S_2$, if there exists
  $s_1 \in S_1$ and $e \in A$ such that $p \rel^{\setminus}
  (s_1,s_2,e,n)$, then $(p,s_2) \notin \rel_2$.
  \begin{itemize}
  \item {\bf Base Case ($n = 1$).} Let $p,s_1,s_2,e$ such that $p
    \rel^{\setminus}(s_1,s_2,e,1)$. If $e \in B_a(s_1,s_2)\cup
    B_b(s_1,s_2) \cup B_d(s_1,s_2)$, then by construction there is an
    $e$ transition in either $P$ or $N_2$ that cannot be matched by the
    other. Thus $(p,s_2)\notin \rel_2$. The same is verified if $e \in
    B_e(s_1,s_2)$ and there is no distribution $\mu_P \in Dist(S_P)$
    such that $L_P(p,e,\mu_P) = \top$.

    Otherwise, $e \in B_e(s_1,s_2) \cup B_c(s_1,s_2) \cup B_f(s_1,s_2)$
    and there exists $\mu_P \in Dist(S_P)$ such that $L_P(p,e,\mu_P) =
    \top$. Let $\phi_1 \in C(S_1)$ and $\phi_2 \in C(S_2)$ be the
    corresponding constraints in $N_1$ and $N_2$. Consider the
    corresponding constraint $\phi_{12}^{B,1} \in C(S)$. By
    $\rel^{\setminus}$, there exists $\mu \in Sat(\phi_{12}^{B,1})$ such
    that $\mu_P \leqbox_{\rel^{\setminus}} \mu$. By construction of
    $\phi_{12}^{B,1}$, we know that either (3.a) there exists $(s'_1,
    \bot, \epsilon, 1)$ such that $\mu(s'_1, \bot, \epsilon, 1) >0$ or
    (3.b) the distribution 
\[\mu_2 : s'_2 \mapsto \sum_{c \in
      A\cup\{\epsilon\}, s'_1 \in S_1, k'\ge 1} \mu(s'_1,s'_2,c, k')\]
    does not satisfy $\phi_2$. If there exists $(s'_1, \bot, \epsilon,
    1)$ such that $\mu(s'_1, \bot, \epsilon, 1) >0$, then there exists
    $p' \in S_P$ such that $\mu_P(p') >0$ and $\textsf{succ}_{s_2,e}(p')
    = \emptyset$. Thus there cannot exists $\mu'_2 \in Sat(\phi_2)$ such
    that $\mu_P \leqbox_{\rel_2} \mu'_2$. Otherwise, by determinism of
    $N_2$, we know that the only possible correspondence function for
    $\mu_P$ and $\rel_2$ is $\delta_2 : S_P \rightarrow (S_2 \rightarrow
    [0,1])$ such that $\delta_2(p')(s'_2) = 1$ if $s'_2 =
    \textsf{succ}_{s_2,e}(p')$ and $0$ otherwise. By construction, we
    have $\mu_P \delta_2 = \mu_2$ and thus there is no distribution
    $\mu_2' \in Sat(\phi_2)$ such that $\mu_P \leqbox_{\rel_2}
    \mu'_2$. Consequently, $(p,s_2) \notin \rel_2$.

  \item {\bf Induction.} Let $1 < n \le K$ and assume that for all $k <
    n$, for all $p' \in S_P, s'_2 \in S_2$, whenever there exists $s'_1
    \in S_1$ and $e \in A$ such that $p'
    \rel^{\setminus}(s'_1,s'_2,e,k)$, we have $(p',s'_2) \notin
    \rel_2$. Let $p,s_1,s_2,e$ such that $p
    \rel^{\setminus}(s_1,s_2,e,n)$. If $e \in B_a(s_1,s_2)\cup
    B_b(s_1,s_2) \cup B_d(s_1,s_2)$, then by construction there is an
    $e$ transition in either $P$ or $N_2$ that cannot be matched by the
    other. Thus $(p,s_2)\notin \rel_2$. The same is verified if $e \in
    B_e(s_1,s_2)$ and there is no distribution $\mu_P \in Dist(S_P)$
    such that $L_P(p,e,\mu_P) = \top$. Else, $e \in B_e(s_1,s_2) \cup
    B_c(s_1,s_2) \cup B_f(s_1,s_2)$ and there exists $\mu_P \in
    Dist(S_P)$ such that $L_P(p,e,\mu_P) = \top$. Let $\phi_1 \in
    C(S_1)$ and $\phi_2 \in C(S_2)$ be the corresponding constraints in
    $N_1$ and $N_2$.

    Consider the corresponding constraint $\phi^{B,n}_{12} \in C(S)$. By
    $\rel^{\setminus}$, there exists $\mu \in Sat(\phi^{B,n}_{12})$ such
    that $\mu_P \leqbox_{\rel^{\setminus}} \mu$.  By construction of
    $\phi^{B,n}_{12}$, we know that either (3.a) there exists $(s'_1,
    \bot, c, 1)$ such that $\mu(s'_1, \bot, c, 1) >0$ or (3.b) the
    distribution 
\[\mu_2 : s'_2 \mapsto \sum_{c \in A\cup\{\epsilon\},
      s'_1 \in S_1, k'\ge 1} \mu(s'_1,s'_2,c, k')\]
    does not satisfy
    $\phi_2$, or (3.c) there exists $s'_1 \in S_1$, $s'_2 \in S_2$, $c
    \ne \epsilon$ and $k < n$ such that $\mu(s'_1,s'_2,c, k)>0$. If case
    (3.a) or (3.b) holds, then as in the base case, there is no
    distribution $\mu_2' \in Sat(\phi_2)$ such that $\mu_P
    \leqbox_{\rel_2} \mu'_2$.  Otherwise, if (3.c) holds, then there
    exists $p' \in S_P$ such that $\mu_P(p') >0$ and $p'
    \rel^{\setminus} (s'_1,s'_2,c,k)$. By induction, we thus know that
    $(p',s'_2) \notin \rel_2$ and by construction and determinism of
    $N_2$, we have that $\textsf{succ}_{s_2,e}(p') = \{s'_2\}$. Thus
    there is no distribution $\mu_2' \in Sat(\phi_2)$ such that $\mu_P
    \leqbox_{\rel_2} \mu'_2$.  Consequently, $(p,s_2) \notin \rel_2$.
  \end{itemize}

  \noindent By hypothesis, we have $s_0^P \rel^{\setminus} (s_0^1,s_0^2,f,K)$. As
  a consequence, we have that $(s_0^P, s_0^2) \notin \rel_2$, implying
  that $P \not \sat N_2$.

  \medskip
  \noindent {\bf 2.} We now prove that for all PA $P \in
  \impl{N_1}\setminus \impl{N_2}$, there exists $K \in \mathbb{N}$ such
  that $P \in \impl{N_1 \setminus^{K} N_2}$.  If $V_1(s_0^1) \ne
  V_2(s_0^2)$, then for all $K\in \mathbb{N}$, we have $N_1
  \setminus^{K} N_2 = N_1$ and the result holds.

  Otherwise, assume that $(s_0^1, s_0^2)$ is in case 3. Let $P = (S_P,
  A, L_P, AP, V_P, s_0^P)$ be a PA such that $P \sat N_1 $ and $P \not
  \sat N_2$. Let $\rel_1$ be the satisfaction relation witnessing $P
  \sat N_1$ and $\rel_2$ be the maximal satisfaction relation between
  $P$ and $N_2$. Assume that $\rel_2$ is computed as described in
  Section~\ref{sec:algo}. Let $\ind_{\rel_2}$ be the associated index
  function and let $K$ be the minimal index such that ${\rel_2}_{K} =
  \rel_2$. We show that $P \sat N_1 \setminus^{K} N_2$. Let $N_1
  \setminus^{K} N_2 = (S,A,L,AP,V,S_0)$ be defined as in
  Section~\ref{sec:under-approx}.

  Let $\rel^{\setminus} \subseteq S_P \times S_2$ be the relation such
  that
  \begin{equation*}
    p \rel^{\setminus} (s_1,s_2,e, k) \iff \left\{ 
      \begin{array}{cl}
        &  (p \rel_1 s_1) \text{ and } (s_2 = \bot) \text{ and } (e =
        \varepsilon) \text{ and } (k = 1) \\
        \text{or} & \left \{ 
          \begin{array}{l}
            (p \rel_1 s_1) \text{ and } (p,s_2) \text{ in case 1 or 2 and }
            (e = \varepsilon)\\
            \qquad \qquad \qquad  \text{ and } (k = 1)
          \end{array} \right. \\ 
        \text{or} & \left \{ 
          \begin{array}{l}
            (p \rel_1 s_1) \text{ and } (p,s_2) \text{ in case 3 and }
            (e \in \breaking(p, s_2)) \\
            \qquad \qquad \qquad \text{ and } (k = \ind_{\rel_2}(p,s_2) +
            1) 
          \end{array} \right.
      \end{array} \right.
  \end{equation*}
  Observe that whenever $(p,s_2)$ is in case 3, we know that
  $\ind_{\rel_2}(p,s_2) < K$, thus $\ind_{\rel_2}(p,s_2) + 1 \le K$.

  We prove that $\rel^{\setminus}$ is a satisfaction relation. Let $p
  \rel^{\setminus} (s_1,s_2,e,k)$.  If $s_2 = \bot$ or $e =
  \varepsilon$, then since $p \rel_1 s_1$, $\rel^{\setminus}$ satisfies
  the axioms of a satisfaction relation by construction.

  Else we have $s_2 \in S_2$ and $e \ne \epsilon$, thus, by definition
  of $\rel^{\setminus}$, we know that $(p,s_2)$ is in case $3$. The rest
  of the proof is almost identical to the proof of
  Theorem~\ref{th:over-diff}. In the following, we report to this proof
  and only highlight the differences.
  \begin{itemize}
  \item By construction, we have $V_P(p) \in V_1(s_1) =
    V((s_1,s_2,e,k))$.

  \item Let $a \in A$ and $\mu_P \in Dist(S_P)$ such that
    $L_P(p,a,\mu_P) = \top$. There are several cases.
    \begin{itemize}
    \item If $a \ne e$, or $a = e \in B_a(p,s_2)$, the proof is
      identical to the proof of Theorem~\ref{th:over-diff}.

    \item Else, we necessarily have $a = e \in B_c(p,s_2) \cup
      B_f(p,s_2)$. Observe that, by construction, $B_c(p,s_2) \subseteq
      B_c(s_1,s_2)$ and $B_f(p,s_2) \subseteq B_f(s_1,s_2)$. Since $p
      \rel_1 s_1$, there exists $\phi_1 \in C(S_1)$ such that
      $L_1(s_1,e,\phi_1) \ne \bot$ and there exists $\mu_1 \in
      Sat(\phi_1)$ and a correspondence function $\delta_1 : S_P
      \rightarrow (S_1 \rightarrow [0,1])$ such that $\mu_P
      \leqbox_{\rel_1}^{\delta_1} \mu_1$.

      Moreover, by construction of $N_1 \setminus^{K} N_2$, we know that
      the constraint $\phi_{12}^{B,k}$ is such that
      $L((s_1,s_2,e,k),e,\phi_{12}^{B,k}) = \top$.

      We now prove that there exists $\mu \in Sat(\phi_{12}^{B,k})$ such
      that $\mu_P \leqbox_{\rel^{\setminus}} \mu$. Consider the function
      $\delta : S_P \rightarrow (S \rightarrow [0,1])$ defined as
      follows: Let $p' \in S_P$ such that $\mu_P(p') >0$ and let $s'_1 =
      \textsf{succ}_{s_1,e}(p')$, which exists by $\rel_1$.
      \begin{itemize}
      \item If $\textsf{succ}_{s_2,e}(p') = \emptyset$, then
        $\delta(p')(s'_1,\bot,\epsilon,1) = 1$.

      \item Else, let $s'_2 = \textsf{succ}_{s_2,e}(p')$. Then,
        \begin{itemize}
        \item if $(p',s'_2) \in \rel_2$, then
          $\delta(p')(s'_1,s'_2,\epsilon,1) = 1$.

        \item Else, $(p',s'_2)$ is in case 3 and $\breaking(p',s'_2) \ne
          \emptyset$. In this case, let $c \in \breaking(p',s'_2)$ and
          define $\delta(p',(s'_1,s'_2,c,\ind_{\rel_2}(p',s'_2)+1)) =
          1$. For all other $c' \in A$ and $1 \le k' \le K$, define
          $\delta(p',(s'_1,s'_2,c',k')) = 0$.
        \end{itemize}
      \end{itemize}

      Observe that for all $p' \in S_P$ such that $\mu_P(p') >0$, there
      exists a unique $s' \in S'$ such that $\delta(p')(s') = 1$. Thus
      $\delta$ is a correspondence function.

      We now prove that $\mu = \mu_P \delta \in Sat(\phi_{12}^{B,k})$.
      \begin{enumerate}
      \item Let $(s'_1,s'_2,c,k') \in S$ such that $\mu(s'_1, s'_2,c,k')
        >0$. By construction, there exists $p' \in S_P$ such that
        $\mu_P(p') >0$ and $\delta(p')(s'_1,s'_2,c,k') >0$. Moreover, $c
        \in B(s'_1,s'_2) \cup \{\epsilon\}$, $s'_2 = \bot$ if
        $\textsf{succ}_{s_2,e}(s_1') = \emptyset$ and $s'_2 =
        \textsf{succ}_{s_2,e}(s_1')$ otherwise.

      \item Consider the distribution $\mu'_1:s'_1 \mapsto \sum_{c \in
          A\cup\{\epsilon\}, s'_2 \in S_2\cup \{\bot\}, k' \ge 1}
        \mu(s'_1,s'_2,c,k')$. By determinism (See Lemma 28
        in~\cite{TCS11}), we have that $\delta_1(p')(s'_1) = 1 \iff s'_1
        = \textsf(succ)_{s_1,e}(p')$. As a consequence, we have that
        $\mu'_1 = \mu \delta_1 = \mu_1 \in Sat(\phi_1)$.

      \item Depending on $k$, there are 2 cases.
        \begin{itemize}
        \item If $k > 1$, assume that for all $p' \in S_P$ such that
          $\mu_P(p') >0$, we have $\textsf{succ}_{s_2,e}(p') \ne
          \emptyset$ (the other case being trivial). Since $c \in
          (B_c(p,s_2) \cup B_f(p,s_2)) \cap \breaking(p,s_2)$ by
          $\rel^{\setminus}$, we can apply Lemma~\ref{lem:ind+}.  As a
          consequence, either (2)~$\mu_1^2 : \big( s'_2 \mapsto \sum_{p'
            \in P \st s'_2 = \textsf{succ}_{s_2,e}(p')} \mu_P(p') \big)$
          does not satisfy $\phi_2$, or (3)~there exists $p' \in S_P$
          and $s'_2 \in S_2$ such that $\mu_P(p') >0$, $s'_2 =
          \textsf{succ}_{s_2,e}(p')$ and $\ind_{\rel_2}(p',s'_2) <
          \ind_{\rel_2}(p,s_2)$.

          In the first case (2), consider the distribution $\mu_2$
          defined as follows:
          \begin{equation*}
            \mu_2 : s'_2 \mapsto
            \sum_{c \in A\cup\{\epsilon\}, s'_1 \in S_1, k' \ge 1}
            \mu(s'_1,s'_2,c, k').
          \end{equation*}
          We have the following: for all $s'_2 \in S_2$,
          \begin{align*}
            \hspace*{8em} \mu_2(s'_2) & = \sum_{c \in A\cup\{\epsilon\},
              s'_1 \in S_1, k' \ge 1}
            \mu(s'_1,s'_2,c, k') \\
            & = \sum_{c \in A\cup\{\epsilon\}, s'_1 \in S_1, k' \ge 1\,}
            \sum_{p' \in S_P} \mu_P(p') \delta(p')((s'_1,s'_2,c, k')) \\
            & = \sum_{p' \in S_P} \mu_P(p') \sum_{c \in
              A\cup\{\epsilon\}, s'_1 \in S_1, k' \ge 1}
            \delta(p')((s'_1,s'_2,c, k')) \\
            &
            \begin{array}{r}
              = {\displaystyle \smash[b]{\sum_{p' \in S_P \st s'_2 =
                  \textsf{succ}_{s_2,e}(p')}} \mu_P(p')
                \delta(p')((\textsf{succ}_{s_1,e}(p'),s'_2,c,} \\[.5ex]
              {\displaystyle \ind_{\rel_2}(p',s'_2)))}
            \end{array} \\
            & \hspace{1cm} \text{ for }c\in
            \breaking(p',s'_2)\text{ fixed as above} \\
            & = \sum_{p' \in S_P \st s'_2 =
              \textsf{succ}_{s_2,e}(p')} \mu_P(p') = \mu_1^2(s'_2)
          \end{align*}
          As a consequence, $\mu_2 \notin Sat(\phi_2)$ and $\mu \in
          Sat(\phi_{12}^{B,k})$.

          In the second case (3), we have $\delta(p')((s'_1,s'_2,c,k'))
          >0$ for $s'_1 = \textsf{succ}_{s_1,e}(p')$, $c \in
          \breaking(p',s'_2)$ fixed above, and $k' =
          \ind_{\rel_2}(p',s'_2)+1 < \ind_{\rel_2}(p,s_2)+1 = k$. As a
          consequence, we thus have $\mu(s'_1,s'_2,c,k') >0$ for $k' <
          k$ and $c \ne \epsilon$, thus $\mu \in Sat(\phi_{12}^{B,k})$.

        \item On the other hand, if $k = 1$, then $\ind_{\rel_2}(p,s_2)
          = 0$ and either (1) there exists $p' \in S_P$ such that
          $\mu_P(p') >0$ and $\textsf{succ}_{s_2,e}(p') = \emptyset$, or
          (2) the distribution
          $\mu_1^2 : \big( s'_2 \mapsto \sum_{p' \in P \st s'_2 =
              \textsf{succ}_{s_2,e}(p')} \mu_P(p') \big) \notin
          \phi_2$. In both cases, as above, we can prove that $\mu \in
          Sat(\phi{12}^{B,k}$.
        \end{itemize}
        In both cases, we have $\mu \in Sat(\phi_{12}^{B,k})$.
      \end{enumerate}
    \end{itemize}

    We thus conclude that there exists $\mu \in Sat(\phi_{12}^{B,k})$
    such that $\mu_P \leqbox_{\rel^{\setminus}} \mu$.

  \item Let $a \in A$ and $\phi \in C(S)$ such that
    $L((s_1,s_2,e),a,\phi) = \top$. As in the proof of
    Theorem~\ref{th:over-diff}, there are several cases that all boil
    down to the same arguments as above.
  \end{itemize}

  \noindent Finally, $\rel^{\setminus}$ is a satisfaction relation: Let $c \in
  \breaking_{\rel_2}(s_0^P, s_0^2)$ and consider the relation
  ${\rel^{\setminus}}' = \rel^{\setminus} \cup \{(s_0^P,(s_0^1,
  s_0^2,c,K))\}$. Due to the fact that $K \ge
  \ind_{\rel_2}(s_0^P,s_0^2)$, one can verify that the pair
  $(s_0^P,(s_0^1, s_0^2,c,K))$ also satisfies the axioms of a
  satisfaction relation. The proof is identical to the one presented
  above. As a consequence, ${\rel^{\setminus}}'$ is also a satisfaction
  relation. Moreover, we now have that $(s_0^P, (s_0^1,s_0^2,c,K)) \in
  {\rel^{\setminus}}'$, with $(s_0^1,s_0^2,c,K) \in S_0$, thus $P \sat
  N_1 \setminus^{K} N_2$.
\qed

\section*{Appendix: Proof of Theorem~\ref{th:c-ex}}

\proof[Proof of Theorem~\ref{th:c-ex}]
  Let $N_1=(S_1,A,L_1,AP,V_1,\{s_0^1\})$ and
  $N_2=(S_2,A,L_2,AP,V_2,\{s_0^2\})$ be deterministic APAs in SVNF such
  that $N_1 \not \preceq N_2$. Let $P = (S,A,L,AP,V,s_0)$ be the
  counterexample defined as above. We prove that $P \sat N_1$ and $P
  \not \sat N_2$.

  \medskip
  \noindent {\bf $\mathbf{P \sat N_1}$.} Consider the relation $\rel_s
  \subseteq S \times S_1$ such that $(s_1,s_2) \rel_s s'_1$ iff $s_1 =
  s'_1$. We prove that $\rel_s$ is a satisfaction relation. Let $t =
  (s_1,s_2) \in S$ and consider $(t,s_1) \in \rel_s$.
  \begin{itemize}
  \item By construction, we have $V(s_1,s_2) \subseteq V_1(s_1)$.

  \item Let $a \in A$ and $\phi_1 \in C(S_1$ such that
    $L_1(s_1,a,\phi_1) = \top$. There are several cases.
    \begin{itemize}
    \item If $(s_1,s_2)$ in case $1$ or $2$ or $s_2 = \bot$, then by
      construction there exists $\mu_1^{\bot} \in Dist(S)$ such that
      $L((s_1,s_2),a,\mu_1^{\bot}) = \top$.  By construction, we have
      that there exists $\mu_1 \in Sat(\phi_1)$ such that $\mu_1^{\bot}
      \leqbox_{\rel_s} \mu_1$.

    \item Else, $(s_1,s_2)$ is in case 3 and $B(s_1,s_2) \ne
      \emptyset$. If $a \notin B(s_1,s_2)$, the result follows as
      above. Else, either $a \in B_a(s_1,s_2) \cup B_b(s_1,s_2)$ and the
      result follows again by construction, or $a \in B_c(s_1,s_2) \cup
      B_f(s_1,s_2)$. In this case, there exists a distribution
      $\widehat{\mu_1} \in Dist(S)$ such that
      $L((s_1,s_2),a,\widehat{\mu_1}) = \top$. By construction,
      $\widehat{\mu_1}$ is defined as follows:
      \begin{equation*}
        \widehat{\mu_1}(s'_1,s'_2) = \left\{
          \begin{array}{ll}
            \mu_1(s_1) & \text{ if } s'_2 = \textsf{succ}_{s_2,e}(s'_1) \\
            & \quad \text { or } \textsf{succ}_{s_2,e}(s'_1) = \emptyset
            \text{ and } s'_2 = \bot \\
            0 & \text{ otherwise }
          \end{array} \right.,
      \end{equation*}
      where $\mu_1$ is either the distribution given by
      Lemma~\ref{lem:ind+} if $a \in \breaking(s_1,s_2)$ or an arbitrary
      distribution in $Sat(\phi_1)$. In both cases, $\mu_1 \in
      Sat(\phi_1)$. Consider the function $\delta : S \times S_1
      \rightarrow [0,1]$ such that $\delta((s'_1,s'_2),s''_1) = 1$ if
      $s'_1 = s''_1$ and $0$ otherwise. Using standard techniques, on
      can verify that $\delta$ is a correspondence function and that
      $\widehat{\mu_1} \leqbox_{\rel_s} \mu_1$.
    \end{itemize}

  \item Let $a \in A$ and $\mu \in Dist(S)$ such that
    $L((s_1,s_2),a,\mu) = \top$. By construction of $P$, there must
    exists $\phi_1 \in C(S_1)$ such that $L_1(s_1,a,\phi_1) \ne \bot$
    and $\mu$ is either of the form $\mu_1^{\bot}$ or $\widehat{\mu_1}$
    for some $\mu_1 \in Sat(\phi_1)$. As above, we can prove that in all
    cases, $\mu \leqbox_{\rel_s} \mu_1$.
  \end{itemize}

  Finally $\rel_s$ is a satisfaction relation. Moreover, we have
  $((s_0^1,s_0^2),s_0^1) \in \rel_s$, thus $P \sat N_1$.

  \medskip
  \noindent {\bf $\mathbf{P \not \sat N_2}$.} Let $\rel_s \subseteq S
  \times S_2$ be the maximal satisfaction relation between $P$ and
  $N_2$, and assume that $\rel_s$ is not empty. Let $\rel \subseteq S_1
  \times S_2$ be the maximal refinement relation between $N_1$ and $N_2$
  and let $K$ be the smallest index such that $\rel_K = \rel$. We prove
  that for all $(s_1,s_2) \in S_1 \times S_2$, if $\ind_{\rel}(s_1,s_2)
  < K$, then $((s_1,s_2),s_2) \notin \rel_s$. The proof is done by
  induction on $k = \ind_{\rel}(s_1,s_2)$. Let $(s_1,s_2) \in S_1 \times
  S_2$.
  \begin{itemize}
  \item {\bf Base case.} If $\ind_{\rel}(s_1,s_2) = 0$, then there are
    several cases.
    \begin{itemize}
    \item If $(s_1,s_2)$ in case 2, i.e. $V_1(s_1) \ne V_2(s_2)$. In
      this case, we know that $V((s_1,s_2)) \in V_1(s_1)$. Thus, by SVNF
      of $N_1$ and $N_2$, we have that $V((s_1,s_2)) \notin V_2(s_2)$
      and $((s_1,s_2),s_2) \notin \rel_s$.

    \item Else, if $(s_1,s_2)$ in cases $3.a$ or $3.b$, then there
      exists $a \in A$ and $\mu_1^{\bot} \in Dist(S)$ such that
      $L((s_1,s_2),a,\mu_1^{\bot}) = \top$ and $\forall \phi_2 \in
      C(S_2)$, we have $L_2(s_2,a,\phi_2) = \bot$. As a consequence,
      $((s_1,s_2),s_2) \notin \rel_s$.

    \item Else, if $(s_1,s_2)$ in cases $3.d$ or $3.d$, then there
      exists $a \in A$ and $\phi_2 \in C(S_2)$ such that
      $L_2(s_2,a,\phi_2) = \top$ and for all $\mu \in Dist(S)$, we have
      $L((s_1,s_2),a,\mu) = \bot$. As a consequence, $((s_1,s_2),s_2)
      \notin \rel_s$.

    \item Finally, if $(s_1,s_2)$ in cases $3.c$ or $3.f$, there exists
      $e \in (B_c(s_1,s_2) \cup B_f(s_1,s_2))\cap
      \breaking(s_1,s_2)$. By Lemma~\ref{lem:ind+}, there exists
      constraints $\phi_1$ and $\phi_2$ such that $L_1(s_1,e,\phi_1) \ne
      \bot$ and $L_2(s_2,e,\phi_2) \ne \bot$ and a distribution $\mu_1
      \in Sat(\phi_1)$ such that either
      \begin{enumerate}[label=(\Roman*)]
      \item \label{item:1-1} $\exists s'_1 \in S_1$ such that
        $\mu_1(s'_1) >0$ and $\textsf{succ}_{s_2,e}(s'_1) = \emptyset$,
        
      \item \label{item:1-2} $\mu_1^2 : \left ( s'_2 \mapsto
          \sum_{\{s'_1 \in S_1 \st s'_2 = \textsf{succ}_{s_2,e}(s'_1)\}}
          \mu_1(s'_1) \right ) \notin Sat(\phi_2)$, or

      \item \label{item:1-3} $\exists s'_1\in S_1, s'_2\in S_2$ such
        that $\mu_1(s'_1)>0, s'_2 = \textsf{succ}_{s_2,e}(s'_1)$ and
        $\ind_{\rel}(s'_1,s'_2) < \ind_{\rel}(s_1,s_2)$.
     \end{enumerate}

     By construction, we have $L((s_1,s_2),e,\widehat{\mu_1}) =
     \top$ for $\mu_1$ given above. Since $\ind_{\rel}(s_1,s_2) = 0$,
     case~(\ref{item:1-3}) above is not possible. From
     cases~(\ref{item:1-1}) and~(\ref{item:1-2}), we can deduce that for
     all $\mu_2 \in Sat(\phi_2)$, we have $\widehat{\mu_1} \not
     \leqbox_{\rel_s} \mu_2$. Moreover, by determinism of $N_2$,
     $\phi_2$ is the only constraint such that $L_2(s_2,e,\phi_2) \ne
     \bot$. As a consequence, $((s_1,s_2),s_2) \notin \rel_s$.
   \end{itemize}

 \item {\bf Inductive step.} Let $0 < k < K$ and assume that for all $k'
   < k$ and for all $(s'_1,s_2) \in \S_1 \times S_2$, if
   $\ind_{\rel}(s_1,s_2) = k'$, then $((s_1,s_2),s_2) \notin \rel_s$.
   Assume that $\ind_{\rel}(s_1,s_2) = k$. There are two cases.
   \begin{itemize}
   \item If $(s_1,s_2)$ in cases 2, $3.a$, $3.b$, $3.d$ or $3.d$, the
     same reasoning applies as for the base case. We thus deduce that
     $((s_1,s_2),s_2) \notin \rel_s$.

   \item Otherwise, if $(s_1,s_2)$ in cases $3.c$ or $3.f$, then, as
     above, there exists $e \in (B_c(s_1,s_2) \cup B_f(s_1,s_2))\cap
     \breaking(s_1,s_2)$. By Lemma~\ref{lem:ind+}, there exists
     constraints $\phi_1$ and $\phi_2$ such that $L_1(s_1,e,\phi_1) \ne
     \bot$ and $L_2(s_2,e,\phi_2) \ne \bot$ and a distribution $\mu_1
     \in Sat(\phi_1)$ such that either
     \begin{enumerate}[label=(\Roman*)]
     \item\label{item:2-1} $\exists s'_1 \in S_1$ such that $\mu_1(s'_1)
       >0$ and $\textsf{succ}_{s_2,e}(s'_1) = \emptyset$,

     \item\label{item:2-2} $\mu_1^2 : \left ( s'_2 \mapsto \sum_{\{s'_1
           \in S_1 \st s'_2 = \textsf{succ}_{s_2,e}(s'_1)\}} \mu_1(s'_1)
       \right ) \notin Sat(\phi_2)$, or

     \item\label{item:2-3} $\exists s'_1\in S_1, s'_2\in S_2$ such that
       $\mu_1(s'_1)>0, s'_2 = \textsf{succ}_{s_2,e}(s'_1)$ and
       $\ind_{\rel}(s'_1,s'_2) < \ind_{\rel}(s_1,s_2)$.
     \end{enumerate}

     By construction, we have that $L((s_1,s_2),e,\widehat{\mu_1}) =
     \top$ for $\mu_1$ given above. As above, if cases~(\ref{item:2-1})
     or~(\ref{item:2-2}) apply, then we can deduce that $((s_1,s_2),s_2)
     \notin \rel_s$. If case~(\ref{item:2-3}) applies, then there exists
     $(s'_1,s'_2) \in S$ such that $\widehat{\mu_1}(s'_1,s'_2) >0$,
     $s'_2 = \textsf{succ}_{s_2,e}(s'_1)$ and $\ind_{\rel}(s'_1,s'_2) <
     \ind_{\rel}(s_1,s_2)$. Since $s'_2 = \textsf{succ}_{s_2,e}(s'_1)$,
     then, by determinism of $N_2$, all correspondence functions
     $\delta$ will be such that $\delta((s'_1,s'_2),s'_2) = 1$. However,
     we have that $\ind_{\rel}(s'_1,s'_2) < k$, thus by induction
     $((s'_1,s'_2),s'_2) \notin \rel_s$. As a consequence, we have that
     for all $\mu_2 \in Sat(\phi_2)$, we have $\widehat{\mu_1} \not
     \leqbox_{\rel_s} \mu_2$. We can thus deduce that $((s_1,s_2),s_2)
     \notin \rel_s$.
   \end{itemize}

   \noindent Finally, we know that $\ind_{\rel}(s_0^1,s_0^2) < k$. As a
   consequence, we have $((s_0^1,s_0^2),s_0^2) \notin \rel_s$ and thus
   $P \not \sat N_2$. \qed
 \end{itemize}

\end{document}